\documentclass[11pt,letterpaper]{article}
\pdfoutput=1
\usepackage{soul}
\usepackage{jheppub}
\usepackage{amsfonts}
\usepackage{amsmath}
\allowdisplaybreaks[4]         
\usepackage{amssymb}   
\usepackage{euscript}       
\usepackage{color}        
\usepackage{slashed} 
\usepackage{tensor}        
\usepackage{starfont}
\newcommand{\labell}[1]{\label{#1}} 

\newcommand{\req}[1]{(\ref{#1})} 

\newcommand{\bea}{\begin{eqnarray}}
\newcommand{\eea}{\end{eqnarray}}
\newcommand{\ba}{\begin{eqnarray}}
\usepackage{braket}
\newcommand{\ea}{\end{eqnarray}}

\newcommand{\beq}{\begin{equation}}
\newcommand{\eeq}{\end{equation} }
\newcommand{\beqa}{\begin{eqnarray}}
\DeclareMathOperator{\Tr}{Tr}
\newcommand{\eeqa}{\end{eqnarray}}
\newcommand{\beqar}{\begin{eqnarray*}}
\newcommand{\eeqar}{\end{eqnarray*}}

\newcommand{\rh}{r_{\rm h}}

\newcommand{\be}{\begin{equation}}
\newcommand{\ee}{\end{equation}}
\newcommand{\diff}{\mathrm{d}}

\renewcommand{\req}[1]{Eq.~(\ref{#1})}

\newcommand{\ssc}{\scriptscriptstyle}
\newcommand{\eg}{{\it e.g.,}\ }
\newcommand{\ie}{{\it i.e.,}\ }

\newcommand{\fin}{\chi_0}

\newcommand{\ctt}{C_{\ssc T}}

\newcommand{\myZ}{\zeta_{\mathfrak{D}}}

\arxivnumber{\tt arXiv:20xx.xxxxx}
\title{ \boldmath Partition functions on slightly squashed spheres and flux parameters}

\author[\text{\Zeus}]{Pablo Bueno,}
\author[\text{\Kronos}]{Pablo A. Cano,}
\author[\text{\Apollon}]{Robie A. Hennigar,}
\author[\text{\Zeus},\text{\Hades}]{\\Victor A. Penas}
\author[\text{\Poseidon}]{and Alejandro Ruip\'erez,}

\affiliation[\text{\Zeus}]{
Instituto Balseiro, Centro At\'omico Bariloche.\\ 8400-S.C. de Bariloche, R\'io Negro, Argentina \vspace{0.1cm}}
\affiliation[\text{\Kronos}]{Instituut voor Theoretische Fysica, KU Leuven.\\
Celestijnenlaan 200D, B-3001 Leuven, Belgium \vspace{0.1cm}}
\affiliation[\text{\Apollon}]{Department of Mathematics and Statistics, Memorial University of Newfoundland.\\ St. John's, Newfoundland and Labrador, A1C 5S7, Canada \vspace{0.1cm}}
\affiliation[\text{\Hades}]{G. F\'isica CAB-CNEA and CONICET, Centro At\'omico Bariloche.\\ 8400-S.C. de Bariloche, R\'io Negro, Argentina \vspace{0.1cm}}
\affiliation[\text{\Poseidon}]{Instituto de F\'isica Te\'orica UAM/CSIC.\\ C/Nicol\'as Cabrera, 13-15, C.U. Cantoblanco, 28049 Madrid, Spain}

\date{\today}
\abstract{
We argue that the conjectural relation between the subleading term in the small-squashing expansion of the free energy of general three-dimensional CFTs on squashed spheres and the stress-tensor three-point charge $t_4$ proposed in {\tt \href{https://arxiv.org/abs/1808.02052}{arXiv:1808.02052}}: $F_{\mathbb{S}^3_{\varepsilon}}^{(3)}(0)=\frac{1}{630}\pi^4 C_{ \ssc T} t_4$, holds for an infinite family of holographic higher-curvature theories. Using holographic calculations for quartic and quintic Generalized Quasi-topological gravities and general-order Quasi-topological gravities, we identify an analogous analytic relation between such term and the charges $t_2$ and $t_4$ valid for five-dimensional theories: $F_{\mathbb{S}^5_{\varepsilon}}^{(3)}(0)=\frac{2}{15}\pi^6 C_{ \ssc T} \left[1+\frac{3}{40} t_2+\frac{23}{630} t_4\right]$. We test both conjectures using new analytic and numerical results for conformally-coupled scalars and free fermions, finding perfect agreement. 
 }

\begin{document} 
\maketitle
\flushbottom

\section{Introduction}
\label{sec:Introduction}

The study of conformal field theories (CFTs) on Euclidean manifolds has proven to be a remarkable source of structural information about such theories. A paradigmatic example corresponds to the free energy of CFTs on spherical backgrounds, which plays a central r\^ole in establishing the monotonicity of renormalization group flows in various dimensions \cite{Klebanov:2011gs,CHM,Casini:2012ei,Myers:2010tj,Myers:2010xs,Pufu:2016zxm,Komargodski:2011vj}. 

On general grounds, the metrics of the corresponding manifolds can be understood as background fields which couple to the stress-energy tensor of the theory, $T_{\mu\nu}$. In particular, the effect of small deformations of the background metric on the partition function is controlled by integrals of various expectation values involving such operator. Since $T_{\mu\nu}$  is defined for every CFT, such deformations are susceptible of having a universal nature ---and they often do \cite{Bobev:2017asb,Fischetti:2017sut,Fischetti:2018shp,Cheamsawat:2018wkr,Bueno:2018yzo,Closset:2012ru}.

In this paper we are interested in the free energy $F_{\mathbb{S}_{\varepsilon}^d}=-\log |Z_{\mathbb{S}_{\varepsilon}^d}|$, of a particular class of odd-dimensional backgrounds, usually called ``squashed spheres'', which preserve a SU$(\frac{d+1}{2})\times$U$(1)$ subgroup of the SO$(d+1)$ isometries group of their round counterparts. Just like those, they are Hopf fibrations over the complex projective space $\mathbb{CP}^{k}$, $\mathbb{S}^1\hookrightarrow \mathbb{S}_{\varepsilon}^d\rightarrow \mathbb{CP}^{k}$, where we used the notation $k\equiv (d-1)/2$. The corresponding metrics can be defined as
\begin{equation}\label{squa}
\diff s^2_{\mathbb{S}_{\varepsilon}^d}=\frac{\diff s^2_{\mathbb{CP}^k}}{(d+1)}+(1+\varepsilon)\left[\diff \psi+\frac{A_{\mathbb{CP}^k}}{(d+1)}\right]^2\, ,
\end{equation}
where $\psi\in [0,2\pi)$ is a periodic coordinate parametrizing the $\mathbb{S}^1$, $J=\diff A_{\mathbb{CP}^k}$ is the K\"ahler form on $\mathbb{CP}^k$, and $\diff s^2_{\mathbb{CP}^k}$ is the Einstein metric on $\mathbb{CP}^k$, normalized so that $R_{ij}=g_{ij}$.\footnote{The most familiar case corresponds to $d=3$, for which $\mathbb{CP}^1 \cong \mathbb{S}^2$, and we can use standard spherical coordinates to write $A_{\mathbb{S}^2}=2\cos \theta \diff \phi$ and $\diff s^2_{\mathbb{S}^2}=\diff \theta^2+\sin^2\theta \diff \phi^2$.}
In the expressions above, the parameter $\varepsilon$ measures the degree of squashing of the sphere, the round case corresponding to $\varepsilon=0$. In general, this parameter can take values in the domain $\varepsilon \in (-1,+\infty)$. 

This kind of backgrounds ---which in the $d=3$ case are sometimes called ``Berger spheres'' \cite{ASNSP_1961_3_15_3_179_0}--- have been often considered in the physics literature in many contexts, including: general field-theoretical studies \cite{Dowker:1998pi,Dowker:2015qta,DeFrancia:2000xm,Bobev:2017asb,Bueno:2018yzo,Zoubos:2004qm}, O$(N)$ models and higher-spin theories \cite{Hartnoll:2005yc,Yonge:2006tn} and holographic cosmology \cite{Anninos:2012ft,Conti:2017pqc,Hertog:2017ymy,Hawking:2017wrd}. Special mention deserves their r\^ole in AdS/CFT \cite{Maldacena,Witten,Gubser}, where the bulk solutions  controlling the corresponding semiclassical partition functions for such boundary metrics correspond to a well-known and important family of gravitational solutions, the so-called Euclidean Taub-NUT/bolt metrics  \cite{10.2307/1969567,doi:10.1063/1.1704018,Hawking:1998ct,Chamblin:1998pz,Emparan:1999pm,Mann:1999pc,Zoubos:2002cw,Bobev:2016sap}. They have been also extensively studied for supersymmetric CFTs\footnote{In that context, Supersymmetry demands the introduction of additional background fields besides the metric, which makes the resulting free energies inequivalent from the ones considered in this paper ---see \cite{Bobev:2017asb} for a more detailed discussion concerning this difference.} ---see \eg \cite{Hama:2011ea,Imamura:2011wg,Closset:2012vp,Closset:2012vg,Martelli:2011fu,Martelli:2011fw,Martelli:2012sz,Witczak-Krempa:2015jca,Toldo:2017qsh}.

Our main interest here will be in the case in which the ``squashing-parameter'' is small,  $|\varepsilon|\ll1$, so we can understand the corresponding backgrounds as small deformations of the usual round sphere and consider a perturbative expansion for the free energy around $\varepsilon=0$. As we review in detail in Section \ref{previouss}, such expansion starts at quadratic order in $\varepsilon$, and the corresponding coefficient is controlled by the flat-space stress-tensor two-point function charge $\ctt$ for general CFTs, namely \cite{Bobev:2017asb,Bueno:2018yzo}
\begin{equation}\label{fo2}
F_{\mathbb{S}_{\varepsilon}^d}= F_{\mathbb{S}^d}+\frac{\varepsilon^2}{2} F_{\mathbb{S}_{\varepsilon}^{d}}^{(2)}(0) +\mathcal{O}(\varepsilon^3)\, , \quad \text{where} \quad F_{\mathbb{S}_{\varepsilon}^{d}}^{(2)}(0)=\frac{(-1)^{\frac{(d-1)}{2}}\pi^{d+1} (d-1)^2 }{  2\,  d!}\ctt\, .
\end{equation} 
The goal of this paper is to provide strong evidence for similar universal relations concerning the $\mathcal{O}(\varepsilon^3)$ terms and the stress-tensor three-point function charges $t_2$ and $t_4$ valid for general three- and five-dimensional CFTs. In particular, we will show that the expressions
\begin{equation}\label{conjs}
F^{(3)}_{\mathbb{S}_{\varepsilon}^{3}}(0)=\frac{\pi^4\ctt }{630} t_4\, , \quad F_{\mathbb{S}_\varepsilon^5}^{(3)}(0) = \frac{2 \pi^6 C_T}{15}  \left[1 +  \frac{3}{40}t_2 + \frac{23}{630}t_4\right]\, ,
\end{equation}
hold for infinite families of  holographic  higher-curvature gravities as well as for free fields.\footnote{The strategy of using higher-curvature gravities as computationally tractable holographic toy models able to teach us lessons concerning universal properties of CFTs  has been exploited in various previous works ---see \eg \cite{Myers:2010tj,Myers:2010xs} for results regarding monotonicity theorems or \cite{Bueno1,Bueno2,Mezei:2014zla,Miao:2015dua,Chu:2016tps,Bianchi:2016xvf} for results regarding entanglement-entropy universal terms.  } The very different nature of the holographic and free-field methods utilized makes us confident that they indeed hold for general theories.

The structure of the paper is the following. In Section \ref{2comm} we introduce various previous results and conjectures involving $F_{\mathbb{S}_{\varepsilon}^d}$ for small values of the squashing parameter and its relation to the flat-space stress-tensor two- and three-point functions. We also review how those quantities can be computed for holographic higher-curvature gravities.  In Section \ref{3dGQTG} we show that the $d=3$ version of our conjecture in \req{conjs} is satisfied by an infinite family of holographic CFTs dual to general-order GQT gravities. In Section \ref{5dtheor} we establish our new conjecture for $d=5$ CFTs using holographic results for quartic GQT gravities. Then, we verify its validity for quintic GQT theories and all-order Quasi-topological gravities.  In Section \ref{freefieee} we use a combination of heat-kernel and zeta-function regularization methods to obtain analytic results for various derivatives of  $F_{\mathbb{S}_{\varepsilon}^3}$ and $F_{\mathbb{S}_{\varepsilon}^5}$ at $\varepsilon=0$ in the case of conformally-coupled scalars ($d=3,5$) and free fermions ($d=3$). We use those to perform exact verifications of the respective conjectures. We conclude in Section \ref{discuss}. In Appendix \ref{hFs}, inspired by results obtained for general GQT gravities in Section \ref{3dGQTG},  we analyze a possible general-CFT relation between the scaling dimension of twist operators and  $F_{\mathbb{S}_{\varepsilon}^3}$, showing that it fails for free fermions. Another plausible relation, in this case between the second derivative of the characteristic function determining the vacua of a given higher-curvature theory and the stress-tensor parameters $t_2$ and $t_4$, conjectured in \cite{Bueno:2018yzo}, is explored in Appendix \ref{upsit2t4} and argued to be false in general. Finally, Appendix \ref{moreQuintic}  contains some additional details concerning the quintic GQT theories used in Section \ref{5dtheor}.


 

\subsubsection*{Note on conventions}
We use latin indices $a,b,\dots$, for bulk tensors and greek indices $\mu,\nu,\dots$, for boundary/CFT tensors, respectively.  $d$ is the spacetime dimensionality of CFTs whereas gravity theories are defined in $(d+1)$ dimensions. $L$ is the action length scale associated to the cosmological constant. The radius of generic AdS$_{(d+1)}$ spaces is denoted by $L/\sqrt{\chi}$. When the background is a solution of the corresponding theory, we replace $\chi$ by $\chi_0$. In the notation of some previous related papers like \cite{Buchel:2009sk,Myers:2010jv,Bueno:2018xqc}, $\chi_0\equiv f_{\infty}$. Our conventions for the holographic charges $\ctt$, $t_2$ and $t_4$ match \eg those of \cite{Myers:2010tj,Buchel:2009sk}. Also, we use the notation $F^{(k)}_{\mathbb{S}_{\varepsilon}^d}(0)\equiv \diff ^k F_{\mathbb{S}_{\varepsilon}^d}/\diff \varepsilon^k|_{\varepsilon=0}$ for the $k$-th derivative of the free energy at $\varepsilon=0$. In order to avoid confusion with the twist-operators scaling-dimension, which we denote by $h_q$, we use $\Upsilon(\chi_0)$ instead of $h(f_{\infty})$ (used \eg in \cite{Bueno:2018yzo}) to denote the characteristic polynomial which determines the AdS vacua of a given bulk theory. 

\section{ CFTs on slightly squashed spheres}\label{2comm}
In this section we start by reviewing some previous results and conjectures concerning the Euclidean partition function of odd-dimensional CFTs on slightly squashed spheres. In the second part we explain how this quantity is computed for holographic theories dual to higher-curvature gravities of the GQT class. Finally, we also summarize how the stress-tensor three-point function charges $t_2$ and $t_4$ can be computed holographically, as well as their relation to the scaling dimension of twist operators and R\'enyi entropies for spherical regions. The methods and results presented in this section will be often called upon (and they will appear intertwined) throughout the paper, so we have preferred to introduce them here for the sake of clarity, simply referring to them when necessary in the remainder of the paper.\footnote{Besides holographic calculations, later we will also present new results for free scalars and fermions. The field-theoretical methods utilized for those will be introduced in the corresponding section.}

\subsection{Previous general results and conjectures}\label{previouss}
For a CFT on some manifold $\mathcal{M}$ with metric $g_{\mu\nu}$, the partition function and associated free energy are defined as
\begin{align}
	Z=\int \mathcal{D} \Phi\, e^{-I_{ E}[\Phi,g_{\mu\nu}]}\, , \quad  F=-\log |Z|\, ,
\end{align}
where $I_{ E}$ is the Euclidean action, and $\Phi$ schematically represents the dynamical fields of the theory. 

Our interest here will be on background metrics corresponding to the special class of squashed spheres defined in the introduction.
For small values of the squashing parameter, $|\varepsilon|\ll1$, we can consider a power-series expansion of $F_{\mathbb{S}_{\varepsilon}^d}$ around $\varepsilon=0$,
\begin{equation}
F_{\mathbb{S}_{\varepsilon}^d}= F_{\mathbb{S}^d}+\sum_{k=1}\frac{1}{k!} F^{(k)}_{\mathbb{S}_{\varepsilon}^d}(0) \varepsilon^k\, .
\end{equation} 
This kind of expansion can be considered with respect to a more general reference metric $\bar g_{\mu\nu}$ by setting $g_{\mu\nu}=\bar g_{\mu\nu}+\varepsilon h_{\mu\nu}$, with $|\varepsilon|\ll1$. Some general results can  in fact be obtained without imposing an explicit form for the perturbation $h_{\mu\nu}$. In particular, assuming $\mathcal{M}_{\bar g_{\mu\nu}}$ to be conformally flat, it can be shown that  \cite{Bobev:2017asb}
\begin{equation}
F^{(1)}(0)=0\, , 
\end{equation}
\ie conformally flat manifolds locally extremize their free energy. Similarly, one can show that the leading non-vanishing contribution is given by \cite{Bobev:2017asb}
 \begin{equation}
F^{(2)}(0)=\gamma(\bar{g}_{\mu\nu},h_{\mu\nu},d)\, \ctt  \, ,
\end{equation}
where $\gamma(\bar{g}_{\mu\nu},h_{\mu\nu},d)$ is a function of the background metric $\bar{g}_{\mu\nu}$, the metric perturbation $h_{\mu\nu}$ and the spacetime dimension ---see \cite{Bobev:2017asb} or \cite{Fischetti:2017sut} for the explicit expression. The function $\gamma(\bar{g}_{\mu\nu},h_{\mu\nu},d)$ is a theory-independent quantity, fully determined by the geometry under consideration. All theory-dependent information contained in $F^{(2)}(0)$ appears through $\ctt$. This is the real and positive quantity ---for unitary CFTs--- which characterizes the flat space two-point function of the stress-energy tensor. Namely,  for general CFTs one has
 \cite{Osborn:1993cr}
\begin{align} \label{t2p}
\braket{ T_{\mu \nu} (x)\, T_{\rho \sigma}(0) }_{\mathbb{R}^d}=\frac{\ctt}{|x|^{2d}}\,\mathcal{I}_{\mu\nu,\rho \sigma}(x)\, ,
\end{align} 
where $\mathcal{I}_{\mu\nu,\rho \sigma}(x)$ is a fixed tensorial structure.\footnote{Its explicit form is 
\begin{equation}\label{Iabcddef}
\mathcal{I}_{\mu\nu,\rho \sigma}(x)\equiv\frac{1}{2}\left[I_{\mu\rho}(x)I_{\nu \sigma}(x)+I_{\mu \sigma}(x)I_{\nu \rho}(x)\right]-\frac{\delta_{\mu\nu}\delta_{\rho\sigma}}{d}\, , \quad \text{where} \quad I_{\mu\rho}(x)\equiv\delta_{\mu\rho}-\frac{2 x_{\mu}x_{\rho}}{|x|^2}\, .
\end{equation}
}

In the case of a slightly squashed sphere of the form \eqref{squa}, $\gamma(\bar{g}_{\mu\nu},h_{\mu\nu},d)$ was computed explicitly using general field-theory techniques for $d=3$ and $d=5$ in \cite{Bobev:2017asb}, the result being
\begin{equation} \label{f20}
F^{(2)}_{\mathbb{S}_{\varepsilon}^3}(0)=-\frac{\pi^4}{3}\ctt\, , \quad \quad F^{(2)}_{\mathbb{S}_{\varepsilon}^5}(0)=+\frac{\pi^6}{15}\ctt\, .
\end{equation}
These were later generalized to arbitrary dimensions in \cite{Bueno:2018yzo} using holographic results to produce the expression in \req{fo2}.
In the general analysis of \cite{Bobev:2017asb}, it was also shown that $F^{(3)}(0)$ was controlled by certain geometry-dependent integrals involving the two- and three-point functions of the stress-tensor, $\braket{ T_{\mu \nu} (x)\, T_{\rho \sigma}(y) }_{\mathbb{R}^d}$, $\braket{ T_{\mu \nu} (x)\, T_{\rho \sigma}(y) T_{\gamma \delta}(z) }_{\mathbb{R}^d}$, as well as an additional term of the form  $\langle \delta T_{\mu\nu}(x)/ \delta g^{\rho \sigma}(y) T_{\alpha\beta}(z)\rangle_{\mathcal{M}}$. Similarly to the two-point function, the stress-tensor three-point function tensorial structure is completely fixed by conformal symmetry for $d$-dimensional CFTs up to three theory-dependent numbers, one of which can be identified with $\ctt$. The other two are customarily denoted by $t_2$ and $t_4$ using the notation of \cite{Hofman:2008ar} ---see Section \ref{3pff}. For parity-preserving CFTs in $d=3$, $t_2$ is absent, and the three-point function is fully controlled by $\ctt$ and $t_4$ alone. 

The presence of the third term described in the  previous paragraph, along with the technical complication associated with the 
general field-theoretical evaluation of the contribution associated with the three-point function, left open the question of whether  
$F_{\mathbb{S}_{\varepsilon}^{d}}^{(3)}(0)$ is fully controlled by some universal combination of $\ctt$, $t_2$ and $t_4$ for general CFTs. This question was partially addressed in \cite{Bueno:2018yzo} using holographic techniques. In that context \cite{Maldacena,Witten,Gubser}, the semiclassical partition function corresponding to a set of boundary conditions is dominated by the $(d+1)$-dimensional bulk geometry with the smallest Euclidean on-shell action compatible with such boundary conditions. The relevant geometries in the case of squashed-sphere boundary metrics of the form \eqref{squa} are those of the Euclidean AdS-Taub-NUT/bolt family \cite{Hawking:1998ct,Chamblin:1998pz,Emparan:1999pm}. These are characterized by a parameter, $n$, called ``NUT charge'' which, by comparing the  boundary metric with \req{squa}, can be related to the squashing parameter. For all holographic theories considered in the present paper, the explicit identification is given by 
\begin{equation}\label{n-epsilon}
\frac{n^2}{\tilde L^2} =\frac{(1+\varepsilon)}{(d+1) }\, ,
\end{equation}
where $\tilde L$ is the AdS radius of the bulk geometry. Using holographic techniques \cite{Emparan:1999pm, Mann:1999pc, Balasubramanian:1999re, Brihaye:2008xu,Bueno:2018xqc}, we can access $F_{\mathbb{S}_{\varepsilon}^d}$ for theories defined by their bulk duals through the evaluation of the regularized Euclidean on-shell action of the corresponding AdS-Taub-NUT solution. For small values of the squashing parameter, NUT geometries typically dominate over their bolt counterparts. This is what happens, for instance, in the case of Einstein gravity, for which the exact result for $F_{\mathbb{S}_{\varepsilon}^d}$ ---valid for\footnote{See \eg \cite{Bobev:2016sap} for related discussions.} $\varepsilon > -(3+\sqrt{3})/(3\sqrt{3})\simeq -0.9107$--- produced by its AdS-Taub-NUT solution is given by \cite{Bobev:2017asb}
\begin{equation}
F^{\rm E}_{\mathbb{S}_{\varepsilon}^d}= F^{\rm E}_{\mathbb{S}^d}+\frac{1}{2}F^{{\rm E}\, (2)}_{\mathbb{S}_{\varepsilon}^{d}}(0)\varepsilon^2\, ,
\end{equation}
where $F^{\rm E\, (2)}_{\mathbb{S}_{\varepsilon}^{d}}(0)$ naturally agrees with the general-CFT result in \req{fo2}, and \cite{Liu:1998bu,Imbimbo:1999bj,Buchel:2009sk}
\begin{equation}\label{einsteinc}
F^{\rm E}_{\mathbb{S}^d}=(-1)^{\frac{d+1}{2}} \frac{\pi^{d/2}}{4\Gamma(d/2)}\frac{\tilde L^{d-1}}{G}\, , \quad \ctt^{\rm E}=\frac{\Gamma(d+2)}{8(d-1)\Gamma(d/2)\pi^{(d+2)/2}}\frac{\tilde L^{d-1}}{G}\, .
\end{equation}
Therefore, the exact Einstein gravity result for the free energy is actually an order-2 polynomial in $\varepsilon$, so its small-$\varepsilon$ expansion is trivial and it stops at that order. Until recently, no additional AdS-Taub-NUT solutions were known in $d+1=4$ bulk dimensions for any other metric theories of gravity. However, new solutions of that kind have been recently constructed in \cite{NewTaub2} for cubic and quartic higher-order gravities of the so-called Generalized Quasi-topological (GQT) class \cite{PabloPablo,Hennigar:2016gkm,PabloPablo2,Hennigar:2017ego,Ahmed:2017jod,PabloPablo3,Hennigar:2017umz,PabloPablo4,Bueno:2019ltp,Bueno:2019ycr,Quasi,Quasi2,Dehghani:2011vu,Cisterna:2017umf}. Remarkably, the thermodynamic properties of such solutions can be obtained fully analytically ---and nonperturbatively in the higher-order couplings--- in all cases. Using the holographic calculation of $t_4$ performed in \cite{Bueno:2018xqc} along with the free-energy result for the cubic theory ---which is the so-called ``Einsteinian cubic gravity'' \cite{PabloPablo,Hennigar:2016gkm,PabloPablo2}--- it was conjectured that  
\begin{equation}\label{3conj}
F^{(3)}_{\mathbb{S}_{\varepsilon}^{3}}(0)=\frac{\pi^4\ctt }{630} t_4\, ,
\end{equation}
holds for general three-dimensional CFTs \cite{Bueno:2018yzo}. This conjecture was tested using the numerical calculations for a conformally-coupled scalar and a free fermion performed in \cite{Bobev:2017asb}, finding agreement with \req{3conj} in both cases. The very different nature of the holographic and free-field calculations suggests the universal validity of the result and, in particular, that the  $\langle \delta T_{\mu\nu}(x)/ \delta g^{\rho \sigma}(y) T_{\alpha\beta}(z)\rangle_{\mathcal{M}}$ term does not contribute to $F^{(3)}_{\mathbb{S}_{\varepsilon}^{3}}(0)$ ---or, alternatively, that it does so in a universal way in terms of $\ctt$ and $t_4$.  Similarly, it is natural to speculate that a similar relation holds in general $d$ between $F^{(3)}_{\mathbb{S}_{\varepsilon}^{d}}(0)$ and some linear combination of $\ctt$, $\ctt t_2$ and $\ctt t_4$. Below we will provide strong evidence in those directions for $d=3$ and $d=5$ theories.

\subsection{Holographic calculations for GQT gravities}
The goal of this paper is to produce additional evidence in favor of the validity of \req{3conj} for general CFTs in $d=3$, and to generalize it to higher dimensions. In order to do so, we will use the holographic dictionary which, as mentioned above, relates the squashed-sphere partition function of a given holographic theory to the on-shell action of some AdS-Taub-NUT bulk solution. In particular, we will consider certain GQT theories allowing for this kind of solutions. GQT theories are
$(d+1)$-dimensional higher-curvature  modifications of the Einstein-Hilbert action of the form
\be  \label{actionHCG}
I = \int \diff^{(d+1)}x \sqrt{|g|} {\cal L}\, , \quad \text{where} \quad {\cal L} = \frac{1}{16\pi G} \left[\frac{d(d-1)}{L^2} + R + \sum_n \lambda_{n} L^{2(n-1)} {\cal R}_{(n)} \right] .
\ee
Here, we assumed a negative cosmological constant characterized by some length scale $L$, the $\mathcal{R}_{(n)}$ are order-$n$ GQT densities, and the $\lambda_n$ are dimensionless couplings.\footnote{In fact, we often consider the possibility of having several densities at a given order, which entails including an additional sum over $i_n$ running over all densities of order $n$.}

 Before defining what GQT gravities are, let us consider a pure AdS solution of a general theory of the form appearing in \req{actionHCG}. Its Riemann tensor is given by
\be  \label{mss}
R_{ab}^{cd} = - \frac{2\chi}{L^2} \delta_{[a}^{[c} \delta_{b]}^{d]} \, ,
\ee
where we have written the curvature radius $\tilde{L}$ in terms of the action scale $L$ and some other dimensionless quantity $\chi$ as $\tilde{L}^2\equiv L^2/\chi$. Let us denote by $\mathcal{L}(\chi)$  the on-shell Lagrangian resulting from evaluating ${\cal L}$ on a maximally symmetric space (mss) for which \req{mss} holds. In terms of this quantity we define the following ``characteristic function''
\begin{align}\label{upside}
\Upsilon(\chi)\equiv \frac{16 \pi G L^2}{d(d-1)} \left[{\cal L}(\chi) - \frac{2}{(d+1)} \chi {\cal L}'(\chi) \right]\, ,
\end{align}
where ${\cal L}'(\chi)$ should be understood as evaluating the Lagrangian density first, and then taking the derivative of the resulting expression with respect to $\chi$.

As shown in \cite{Aspects}, imposing a maximally symmetric space to be a solution of \req{actionHCG} for a general higher-curvature theory boils down to imposing
\be  \label{polyh}
\Upsilon(\chi_0) =0\, . 
\ee
While \req{polyh} is the condition for a certain mss to be a solution of the corresponding theory, we can also consider $\Upsilon(\chi)$ as defined in terms of $\mathcal{L}(\chi)$ in \req{upside} ``off-shell'', namely, evaluated for some other argument and without imposing such condition. Whenever we are considering a possible vacuum of the theory, we will denote the argument of $\mathcal{L}$, $\Upsilon$ or their derivatives with respect to $\chi$ by ``$\chi_0$''. 
If we take $\mathcal{L}$ to be a linear combination of densities of the form (\ref{actionHCG}), when we evaluate it on a mss satisfying \req{mss}, $\Upsilon(\chi)$ becomes an order-$n$ polynomial of the form
$
\Upsilon(\chi)=1-\chi+\sum_n c_n \lambda_n \chi^n
$
for certain constants $c_n$. A somewhat canonical normalization for the $\mathcal{R}_{(n)}$ consists then in rescaling the densities as $\mathcal{R}_{(n)}\rightarrow \mathcal{R}_{(n)}/c_n$, so that $\Upsilon(\chi)$ takes the form
\be \label{hh}
\Upsilon(\chi)=1-\chi+\sum_n  \lambda_n \chi^n\, .
\ee 
We will be assuming our densities throughout the paper to be normalized in this way whenever possible. Given this normalization of the densities, the on-shell Lagrangian $\mathcal{L}(\chi)$ takes the form\footnote{Notice the $(d+1-2n)$ factor in the denominator of the last term. This is because order-$n$ densities do not contribute to the equations of motion of mss in $d+1=2n$ dimensions.}
\be \label{LL}
\mathcal{L}(\chi)=\frac{d(d-1)}{16\pi G L^2}\left[1-\frac{(d+1)}{(d-1)} \chi +  \sum_n \frac{(d+1) }{(d+1-2n)} \lambda_n \chi^n \right]\, .
\ee

The function $\Upsilon(\chi)$ will play an important r\^ole in our discussion. As a first property, it was shown in \cite{Bueno:2018yzo} that, for general Einstein-like theories,\footnote{By ``Einstein-like theories'' here we mean higher-curvature gravities whose linearized spectrum around general maximally symmetric spaces only includes the standard transverse and traceless graviton of Einstein gravity \cite{Aspects}. } the two-point function charge $\ctt$ for holographic theories dual to this kind of bulk theories is given by
\begin{equation}\label{cte}
\ctt=- \Upsilon'(\chi_0) \ctt^{\rm E}\, ,
\end{equation}
where $\ctt^{\rm E}$ is the Einstein gravity result appearing in \req{einsteinc}.




\subsubsection{GQT NUTs free energies and squashed spheres}

Most of the above discussion is valid for a general action of the form \req{actionHCG}. Let us now restrict ourselves to GQT theories. Their defining property is the following \cite{Hennigar:2017ego,Bueno:2019ltp,Bueno:2019ycr}. Consider a general static and spherically symmetric metric of the form
\be \label{staticsphere}
\diff s^2=-N(r)^2V(r) \diff t^2+V(r)^{-1}\diff r^2+r^2 \diff \Omega_{(d-1)}^2\, ,
\ee
and let $L_{N,V}\equiv \left. \sqrt{-g} {\cal L} \right|_{N,V}$ be the effective Lagrangian resulting from the evaluation of $\mathcal{L}$ on \eqref{staticsphere}. We say the corresponding theory is of the GQT class if the Euler-Lagrange equation of  $V$  associated to $L_{V}\equiv L_{N=1,V}$ is identically satisfied.  In that case, one can set $N(r)=1$, and the corresponding solutions satisfy $g_{tt}g_{rr}=-1$. In general, $V(r)$ turns out to satisfy a second-order differential equation. In some cases, however, this order gets reduced and $V(r)$ is characterized by an algebraic equation of order $n$. Theories satisfying this latter property are called ``Quasi-topological (QT) gravities'' \cite{Quasi2,Quasi,Myers:2010jv,Oliva:2011xu,Oliva:2012zs,Dehghani:2011vu,Dehghani:2013ldu,Cisterna:2017umf}. Naturally, from this perspective, Lovelock theories \cite{Lovelock1,Lovelock2} are in turn a particular subset of QT theories. GQT theories exist in general dimensions and at arbitrarily high orders in curvature \cite{Bueno:2019ycr}, and they have many interesting properties, such as possessing second-order equations of motion when linearized around maximally symmetric backgrounds, or the fact that the thermodynamic properties of their black hole solutions can be computed analytically  ---see \eg \cite{Bueno:2019ltp} for a detailed summary. 

For the purposes of this work, the most relevant aspect is that a certain subset of  GQT theories admit solutions of the AdS-Taub-NUT class which are also characterized by a single function and whose thermodynamic properties can be computed analytically \cite{Clarkson:2002uj,KhodamMohammadi:2008fh,NewTaub2}. As explained in \cite{NewTaub2,Bueno:2018yzo}, the relevant solutions take the general form
\begin{equation}\label{FFnut}
\diff s^2=V_{\mathbb{CP}^k}(r) \left[\diff \tau+ n A_{\mathbb{CP}^k}\right]^2+V_{\mathbb{CP}^k}(r)^{-1}\diff r^2+\left[r^2-n^2\right]\diff s_{\mathbb{CP}^k}^2\, ,
\end{equation}
where $n$ is the NUT charge (not to be confused with the order of the higher-curvature terms). For even $(d+1)$, one can replace $\mathbb{CP}^{\frac{d-1}{2}}$ by any other $(d-1)$-dimensional K\"ahler-Einstein manifold $\mathcal{B}$, and the Taub-NUT/bolt solutions will correspond to U$(1)$ fibrations over $\mathcal{B}$. Similarly to $\psi$ in \req{squa},  $\tau$ is a periodic coordinate parametrizing the U$(1)$, whose period must be fixed to $\beta_{\tau}=2n(d+1)\pi$ in order to eliminate the Dirac-Misner string \cite{Misner:1963fr} associated to $A_{\mathbb{CP}^k}$.
The fact that the solution should be locally asymptotically AdS imposes
$
V_{\mathbb{CP}^k}(r)=r^2/\tilde L^2+\mathcal{O}(1)$
for
$r\rightarrow\infty$. From this, it follows that the boundary metric is indeed conformally equivalent to the squashed-sphere one appearing in \req{squa} with squashing parameter related to the NUT charge through \req{n-epsilon}. In general, there will be a value of $r=r_{\rm \ssc H}$ such that $V(r_{\rm \ssc H})=0$. Whenever $r_{\rm \ssc H}=n$ the solution is called a ``NUT'', whereas for $r_{\rm \ssc H}\equiv r_b >n$ it is a ``bolt''. For both types of solutions, imposing regularity in the bulk fixes 
$
V_{\mathbb{CP}^k}'(r_{\rm \ssc H})=4\pi/\beta_{\tau}
$. In all cases considered here, the relevant free energy is the one corresponding to the NUT solution, since it is the one that dominates the partition function for $|\varepsilon| \ll 1$. In the GQT theories of interest for us, the equations of motion collapse to a single equation for $V_{\mathbb{CP}^k}(r)$ which can be integrated once, producing a second-order equation of the form $\mathcal{E}\left[V_{\mathbb{CP}^k}(r),V'_{\mathbb{CP}^k}(r),V''_{\mathbb{CP}^k}(r),r\right]=C$, where $C$ is an integration constant proportional to the energy of the solution. In all cases, imposing the solution to be locally asymptotically AdS$_{(d+1)}$ as well as regularity in the interior completely determine it ---see \cite{NewTaub2} for numerous explicit examples. 

Now, given some AdS-Taub-NUT solution of this kind for certain GQT gravity, we need to compute the corresponding Euclidean on-shell action in order to access the free energy of the dual CFT on a squashed sphere. The standard way of performing such calculation involves the introduction of generalized versions \cite{Teitelboim:1987zz,Myers:1987yn} of the Gibbons-Hawking boundary term \cite{York:1972sj,Gibbons:1976ue} as well as counterterms \cite{Emparan:1999pm, Mann:1999pc, Balasubramanian:1999re, Brihaye:2008xu} which render the resulting action finite. A simplified method which only requires the Einstein gravity boundary term plus knowledge of $F_{\mathbb{S}^d}$ ---or, equivalently, the quantity customarily denoted $a^*$--- valid for GQTs was introduced in \cite{Bueno:2018xqc} and successfully applied later in \cite{NewTaub2,Mir:2019rik}.  

Interestingly, it has been pointed out in \cite{Bueno:2018yzo} that the free energy of all NUT solutions constructed so far for GQT gravities compatible with the ansatz \eqref{FFnut} can be computed using an auxiliary pure AdS$_{(d+1)}$ with a rescaled radius given by $L \sqrt{(1+\varepsilon)/\chi_0}$. Explicitly, the proposed expression reads
\begin{equation}\label{fee0e}
F_{\mathbb{S}_{\varepsilon}^{d}}=(-1)^{\frac{(d-1)}{2}}\frac{\pi^{\frac{(d+2)}{2}} }{\Gamma\left[\frac{d+2}{2}\right]}\frac{\mathcal{L}\left[\chi_0/(1+\varepsilon)\right] L^{d+1}}{[\chi_0/(1+\varepsilon)]^{\frac{(d+1)}{2}}}\, ,
\end{equation}
where we stress that $\mathcal{L}\left[\chi_0/(1+\varepsilon)\right]$ should be understood as the corresponding GQT Lagrangian evaluated on a pure AdS of the form \req{mss} with $\chi$ replaced by $\chi_0/(1+\varepsilon)$. 

This formula satisfies a number of consistency checks \cite{Bueno:2018yzo}: i) it reduces to the round-sphere result valid for general higher-curvature gravities \cite{Imbimbo:1999bj,Schwimmer:2008yh,Myers:2010tj,Myers:2010xs,Bueno:2018xqc} when we set $\varepsilon=0$; ii) it correctly yields a vanishing result for its first derivative with respect to $\varepsilon$ at $\varepsilon=0$, since $F^{(1)}_{\mathbb{S}_{\varepsilon}^{d}}(0)\propto \Upsilon(\chi_0)=0$; iii) it also produces the right dependence on $\ctt$ for the second derivative appearing in \req{fo2}, as can be easily verified using \req{cte} and \req{einsteinc}. Assuming its validity, we can use \req{LL} to write explicitly
\be \label{LL}
F_{\mathbb{S}_{\varepsilon}^{d}}=\frac{(-1)^{\frac{(d-1)}{2}}\pi^{d/2} }{16 \Gamma\left[\frac{d+2}{2}\right]}\frac{d(d-1) L^{d-1}}{ [\chi_0/(1+\varepsilon)]^{\frac{(d+1)}{2}}G} \left[1-\frac{(d+1)}{(d-1)} \frac{\chi_0}{(1+\varepsilon)} +  \sum_n \frac{(d+1) }{(d+1-2n)}  \frac{\lambda_n \chi_0^n}{(1+\varepsilon)^n}  \right]
\, ,
\ee 
which we conjecture to be valid for general GQT theories admitting AdS-Taub-NUT solutions of the form given by \req{FFnut}. The evidence in favor of this conjecture includes Gauss-Bonnet gravity in general dimensions, cubic and quartic GQTs in $d=3$, as well as a quartic GQT and a quartic QT in  $d=5$. Below, we will provide additional evidence for its validity for general-order GQTs in $d=3$ and up to $n=5$ in $d=5$ as well as for general-dimension and general-order QT gravities.


\subsubsection{Stress tensor three-point function, energy fluxes and twist operators}\label{3pff}
So far, we have reviewed the known general results concerning the free energy of slightly squashed spheres and the way such quantity is computed for holographic GQT gravities. Our plan  is to study possible universal relations between $F^{(3)}_{\mathbb{S}_{\varepsilon}^{d}}(0)$ and the stress-tensor charges $C_T$, $t_2 $ and $t_4$, so let us briefly explain now how the latter can be accessed for holographic higher-curvature gravities.

A standard method for computing $t_2$ and $t_4$ in holographic theories follows from the thought experiment proposed in \cite{Hofman:2008ar}. The idea is to consider an insertion of the stress tensor  $\mathcal O \sim \varepsilon^{ij}T_{ij}$ on the vacuum (for some arbitrary constant polarization tensor $\varepsilon^{ij}$), and then compute the expectation value of the energy flux measured far away from the perturbation in some direction  $\vec n$ in the resulting state.
Using standard coordinates in Minkowski space, so that the metric reads $ds^2=-dt^2+ \delta_{ij}dx^i dx^j$, the energy flux in the direction $\vec n$ is given by
\begin{equation}\label{eq:flux-operator}
\mathcal E\left(\vec n\right)=\lim_{r\to \infty}r^{d-2}\int_{-\infty}^\infty dt\, T^t{}_i \left(t,rn^i\right)n^i \ ,
\end{equation}
where $r^{2}\equiv \delta_{ij} x^i x^j$. For any CFT$_d$ in $d\geq 4$, the expectation value of the energy flux in the excited state $\mathcal{O}\ket{0}$ is given by \cite{Hofman:2008ar,Buchel:2009sk}
\begin{equation}\label{eq:flux}
\langle\mathcal E\left(\vec n\right)\rangle=\frac{E}{4\pi}\left[1+t_2\left(\frac{\varepsilon^*_{ij}\varepsilon_{ik}n^in^k}{\varepsilon^*_{ij}\varepsilon_{ij}}-\frac{1}{d-1}\right)+t_4\left(\frac{|\varepsilon_{ij}n^in^j|^2}{\varepsilon^*_{ij}\varepsilon_{ij}}-\frac{2}{d^2-1}\right)\right]\ ,
\end{equation}
where $E$ is the total energy. Since the tensorial structures appearing in this expression are completely fixed for any CFT, we can extract the values of $t_2$ and $t_4$ for a given theory by evaluating $\langle\mathcal E\left(\vec n\right)\rangle$ and identifying the coefficients proportional to such structures.
Holographically, this amounts to evaluating the corresponding Euclidean action on the following perturbation of $\mathrm{AdS}_{d+1}$\footnote{This method has been used to identify $t_4$ and $t_2$ for holographic theories dual to certain higher-order gravities in $d\geq 4$, including Lovelock \cite{Buchel:2009sk,deBoer:2009gx}, cubic QTG \cite{Myers:2010jv} and general cubic theories \cite{Li:2019auk}.}
\begin{equation}\label{eq:perturbationsAdSD}
\begin{aligned}
ds^2=&\frac{\tilde L^2}{u^2}\left[\delta(y^+)W\left(y^i,u\right)\left(dy^+\right)^2-dy^+dy^-+\sum_{i=1}^{d-2}\left(dy^i\right)^2+du^2\right]\\
&+h_{++} \left(dy^+\right)^2+2h_{+1}dy^+dy^1+2h_{+2}dy^+dy^2+2h_{12}dy^1dy^2 \, ,
\end{aligned}
\end{equation}
where we used coordinates
\begin{equation}
y^+\equiv -\frac{1}{x^0+x^3}\ , \quad y^-\equiv x^0-x^3-\frac{x^ix^i}{x^0+x^3}\ , \quad y^i\equiv \frac{x^i}{x^0+x^3}\ ,
\end{equation}
for $i=1, 2, \dots, d-2$. 

Let us briefly explain this. The metric (\ref{eq:perturbationsAdSD}) represents two different perturbations of $\mathrm{AdS}_{d+1}$. The first line corresponds to a shockwave background which is dual to the flux operator $\mathcal E(\vec n)$. As it turns out, the equations of motion for the metric evaluated on the shockwave ansatz, are exactly the same as for Einstein gravity for a general higher-curvature theory \cite{Horowitz:1999gf}. They read
\begin{equation}\label{eq:WEOM}
\partial_u^2W -\frac{d-1}{u}\partial_u W+\sum_{i=1}^{d-2}\partial_i^2W=0\, .
\end{equation}
One can explicitly check that a solution to this equation is
\begin{equation}\label{eq:solW}
W(y^1,y^2,u)=\frac{W_0\,u^d}{\left(u^2+(y^1-y^1_0)^2+(y^2-y^2_0)^2\right)^{d-1}}\, ,
\end{equation}
where $y^i_0=n^i/(1+n^{d-1})$ and $W_0$ is a normalization constant that plays no r\^ole in the discussion. The second line of (\ref{eq:perturbationsAdSD}) represents the metric perturbation dual to the localized insertions of the energy momentum $\mathcal O$ for the particular polarization chosen, \eg $\varepsilon_{x^1x^2}=1$. The remaining components of the perturbation $h_{++}, h_{+1}$ and $h_{+2}$ must be turned on in order to ensure that the perturbation is transverse $\nabla^\mu h_{\mu\nu}=0$. The transverse condition then imposes 
\begin{equation}\label{eq:transverseconditions}
\partial_- h_{+1}=\frac{1}{2}\partial_2 h_{21}\, , \quad \partial_- h_{+2}=\frac{1}{2}\partial_1 h_{12}\, , \quad \partial_- h_{++}=\frac{1}{2}\left(\partial_1h_{1+}+\partial_2 h_{2+}\right)\, .
\end{equation}
This turns out to be crucial for the calculations, as it typically simplifies rather drastically the equations of motion of the perturbation $h_{\mu\nu}$. Ignoring interaction terms with the shockwave, we have
\begin{equation}\label{eq:phiEOM}
\partial_u^2\phi -\frac{d-1}{u}\partial_u \phi+\sum_{i=1}^{d-2}\partial_i^2\phi-4\partial_+\partial_-\phi=0\ ,
\end{equation}
where $\phi\equiv \tfrac{u^2}{\tilde L^2} h_{12}$. All that remains to compute the flux parameters $t_2$ and $t_4$ is to evaluate the corresponding action on the metric (\ref{eq:perturbationsAdSD}) keeping only terms linear in $W$ and quadratic in $\phi$. After using the transverse conditions (\ref{eq:transverseconditions}), the equations of motion of the shockwave (\ref{eq:WEOM}) and of $\phi$ (\ref{eq:phiEOM}), and several integrations by parts, the piece of the on-shell action of interest for us will take the following form
\begin{equation}
I_E=-\frac{1}{16\pi G}\int d^dy du \frac{\tilde L^{d-1}\delta(y^+)W \,\phi\, \partial_- \phi}{u^{d-1}}\left[k_0+k_2 \,T_2 +k_4 \,T_4\right]\, ,
\end{equation}
where $k_{0}, k_{2}$ and $k_{4}$ are theory-dependent constants and $T_2$ and $T_4$ are functions that depend on $u, W$ and its derivatives\footnote{They are homogeneous functions of degree 0 in $W$, \textit{i.e.}, they are of the form $\sim \partial^2W/W$.} and whose specific form depends on the dimension $d$. 
 These functions, evaluated at the point $u=1, y^i=0$, are proportional to the tensorial structures appearing in front of $t_2$ and $t_4$ in the general expression for the expectation value of the integrated energy flux (\ref{eq:flux}), which will finally allow us to obtain the values of $t_{2}$ and $t_{4}$ for a given higher-curvature theory.

Besides energy fluxes, there exist additional interesting quantities universally connected with $t_2$ and $t_4$. This is the case of R\'enyi entropies $S_q$ for spherical entangling regions and ---related to these--- the scaling dimension of the ``twist''  operators whose expectation value yields the trace of the $q$-th power of the reduced density matrix involved in the definition of $S_q$. More precisely, consider some spatial subregion $V$ and its complement $\bar V$. The $q$-th R\'enyi entropy is defined as \cite{renyi1,renyi1961}
\begin{equation}\label{rr}
S_q(V)=\frac{1}{1-q}\log \Tr \rho_V^q \, , \quad q\geq 0\, ,
\end{equation}
where $\rho_V$ is the partial-trace density matrix obtained integrating over the degrees of freedom in  $\bar V$. 
The trace  $\Tr \rho_V^q$ can be obtained as the expectation value of certain dimension-$(d-2)$ twist operators $\tau_{q}$ defined over $\partial V$ \cite{Calabrese:2004eu,Hung:2011nu,Hung:2014npa,Swingle:2010jz}. This expectation value is computed in the symmetric product of $q$ copies of the theory defined on a single copy of the geometry, $\Tr \rho_V^q=\braket{\tau_q}_q$ ---in contradistinction to the usual replica trick, where one defines the theory in a manifold which involves $q$ different copies of the geometry sewn together at $\partial V$. The leading singularity in the correlator  $\braket{T_{\mu\nu}\tau_q}$ defines the conformal dimension of  $\tau_q$ \cite{Kapustin:2005py,Hung:2011nu,Hung:2014npa}. In particular, if we make an insertion of the stress-tensor near $\partial V$, it can be argued that such correlator is given ---regardless of the geometry of $V$--- by \begin{equation}
\braket{T_{\mu\nu}\tau_q}_q=-\frac{h_q}{2\pi} \frac{c_{\mu\nu}}{y^d}+\text{subleading}\, ,
\end{equation}
where $y$ is the separation between the insertion of $T_{\mu\nu}$ and $\partial  V$, and $c_{\mu\nu}$ is a fixed tensorial structure.

The most relevant aspect for our purposes is that derivatives of $h_q$ and $S_q$ evaluated at $q=1$ produce expressions which are related to correlators of the stress energy tensor. In particular, one finds \cite{Hung:2014npa,Chu:2016tps}
\begin{align}\label{hq1}
\left.\partial_q h_q \right|_{q=1} &= 2 \pi^{\frac{d}{2}+1}\frac{\Gamma(d/2)}{\Gamma(d+2)}\, \ctt \, ,\\ \label{hq2t}
\left.\partial^2_q h_q \right|_{q=1} &= - \frac{2 \pi^{1 + d/2} \Gamma(d/2) \ctt}{(d-1)^3 d (d+1)\Gamma(d+3)} \bigg[ 
d(2 d^5 - 9 d^3 + 2 d^2 + 7 d - 2) 
\\ \notag
&+ (d-2)(d-3)(d+1)(d+2)(2d-1)t_2 + (d-2)(7d^3 - 19 d^2 - 8 d + 8) t_4
\bigg]\, ,
\end{align}
with similar expressions holding for $\left.\partial_q S_q \right|_{q=1}$ and $\left.\partial^2_q S_q \right|_{q=1}$ \cite{Perlmutter:2013gua,Lee:2014zaa}.
 Evaluated for $d=3$ and $d=5$, respectively, $\left.\partial^2_q h_q \right|_{q=1}$ is given by
 \begin{equation}\label{hq2eval}
 \left.\partial^2_q h_q \right|^{d=3}_{q=1}=-\frac{\pi^4 \ctt}{5760} [420+t_4] \, , \quad  \left.\partial^2_q h_q \right|^{d=5}_{q=1}= -\frac{3\pi^4\ctt}{640}\left[\frac{31}{36}+\frac{3}{40}t_2+\frac{23}{630} t_4 \right]\, .
 \end{equation}

Holographically, both $S_q$ and $h_q$ are in general much simpler to compute than the expectation value of the energy flux $\braket{\mathcal{E}(\vec{n})}$ considered above. Indeed, both quantities  can be obtained as \cite{CHM,Hung:2011nu,Hung:2014npa}
  \begin{align}\label{sex}
S_q=\frac{q}{(q-1)T_0}\left[S(x)T(x)-E(x)\right]\big|^1_{x_q}\, , \quad
h_q=\frac{-2\pi R q}{(d-1)V_{\mathbb{H}^{d-1}}}\left(E(x_q)-E(1)\right)\, ,
\end{align}
where $T$, $S$ and $E$ stand, respectively, for the temperature, thermal entropy and energy of the hyperbolic AdS black hole of the bulk theory considered, $R$ is the radius of the hyperbolic space, and we defined $x\equiv \rh \sqrt{\chi_0}/ L$. On general grounds, one has $T(1)=T_0\equiv 1/(2\pi R)$, while $x_q$ is defined as the real solution to the equation $T(x_q)=T_0/q$ which reduces to the Einstein gravity one in the appropriate limit. This means that, given a bulk theory with a hyperbolic-horizon static black hole solution whose thermodynamic properties we can compute, the particular linear combination of   $t_2$ and $t_4$ appearing in \req{hq2t} can be obtained using that equation after evaluating $h_q$ using \req{sex}. This is particularly useful in $d=3$. In that case, $t_2$ is absent, and $t_4$ can be obtained from $\left.\partial^2_q h_q \right|_{q=1} $ ---this was the method used in \cite{Bueno:2018xqc} for Einsteinian cubic gravity.


This concludes our extended summary of previous general results and conjectures regarding the free energy of slightly squashed spheres as well as of holographic methods for the computation of such quantity and of the flux parameters $t_2$ and $t_4$.

\section{Three-dimensional holographic CFTs}\label{3dGQTG}
In this section we compute the thermodynamic properties of hyperbolic black holes for general-order GQT gravities in $d=3$. Using this, we extract $t_4$ from the scaling dimension of twist operators. Then, we show that the original conjecture (\ref{3conj}) relating the subleading term in the slightly squashed-sphere expansion to $t_4$ holds for this infinite family of holographic higher-curvature gravities.
\subsection{General GQT theories}\label{gqt3}
Recently, some of us have shown that GQT gravities exist at all orders in curvature by providing both recursive and explicit all-order formulas \cite{Bueno:2019ycr}. On the other hand, it is known that not all GQTGs admit single-function Taub-NUT solutions \cite{NewTaub2} and, at this point, we do not possess a full all-order characterization of  those theories, for which we expect the master free-energy formula \req{fee0e} to hold. Nevertheless, we do know that all such theories are a subset of the GQT class, and in addition, we know that $D=4$ GQTGs modify in a unique way the static black hole solutions at every order in curvature. Therefore, the thermodynamic properties of static black holes in theories admitting single-function Taub-NUTs are the same as those of ordinary GQTGs. We will use this fact to compute the entropy and temperature of hyperbolic black holes in theories allowing for single-function Taub-NUTs, from where we will extract the scaling dimension of twist operators, $h_q$, which we will use to obtain $t_4$ for those theories using \req{hq2t}. 

Let us then consider a general GQT theory 
 involving an infinite number of higher-derivative terms,
\begin{equation}\label{genGQTG3d}
I=\frac{1}{16\pi G}\int \diff^4x\sqrt{|g|}\left[\frac{6}{L^2}+R+\sum_{n=3}^{\infty}\lambda_{n}L^{2n-2}\mathcal{R}_{(n)}\right]\, .
\end{equation}
For $n=3$ and $n=4$ we can choose \cite{NewTaub2}
\begin{align}
\mathcal{R}_{(3)}= &\,-\frac{1}{8}\left[12 R_{a\ b}^{\ c \ d}R_{c\ d}^{\ e \ f}R_{e\ f}^{\ a \ b}+R_{ab}^{cd}R_{cd}^{ef}R_{ef}^{ab}-12R_{abcd}R^{ac}R^{bd}+8R_{a}^{b}R_{b}^{c}R_{c}^{a} \right] \, ,
\\
\mathcal{R}_{(4)}=&-\frac{1}{16}\left[-44R^{abcd }R_{ab }^{\ \ ef }R_{c\ e}^{\ g \ h}R_{d g f h}-5 R^{abcd }R_{ab }^{\ \ ef }R_{ce}^{\ \ gh}R_{d f g h}\right.\\ \notag & \quad \quad \quad \left.+5 R^{abcd }R_{abc }^{\ \ \ \ e}R_{f g h d}R^{f g h}_{\ \ \ \ e} +24 R^{ab }R^{cd ef }R_{c\ e a}^{\ g}R_{d g f b}\right]\, .
\end{align}
Additional explicit densities for $n=3,\dots,8$ can be found in \cite{Arciniega:2018tnn}.
For general $n$, the densities $\mathcal{R}_{(n)}$ are such that they allow for single-function Taub-NUT solutions, whose existence at arbitrary $n$ is assumed. While we do not have a closed expression for them for general $n$, we know that when evaluated on a spherical/planar/hyperbolic black hole ansatz, they are equivalent to the densities constructed in \cite{Bueno:2019ycr}. Therefore, they produce the same on-shell actions, equations of motion, and so on. 

A general hyperbolic black hole ansatz is given by
\begin{equation}\label{eq:hypBH3d1}
\diff s^2=-N(r)^2V(r)\diff t^2+\frac{\diff r^2}{V(r)}+r^2 \diff \Xi^2\, , \quad \text{where} \quad  \diff \Xi^2=\diff \theta^2+\sinh^2\theta \diff \phi^2\, ,
\end{equation}
represents the metric of the unit hyperbolic space, 
and where in principle $N(r)$ and $V(r)$ are two independent functions. 
The equations of motion of \eqref{genGQTG3d} evaluated on the metric \eqref{eq:hypBH3d1} were computed in \cite{Bueno:2019ycr}, where it was found that they are solved by $N(r)=$ constant, while $V(r)$ satisfies an equation which can be most conveniently written by defining 
\begin{equation}\label{Vdef}
V(r)\equiv \frac{r^2}{L^2}f(r)-1\, .
\end{equation}
In terms of $f(r)$, it reads
\begin{align}\label{eq:V3dEq} 
&r^3(1-f)+\sum_{n=3}^{\infty} \lambda_{n}\left(f+\frac{r f'}{2r}\right)^{n-3}\Bigg[f^3 r^3-\frac{ f^2 f'}{2} (n-3) r^4-\frac{f'^3}{8}(n-1) r^6\\ \notag
&+ \frac{f'^2 r^3}{8}\left[ f \left(6-7 n+3 n^2\right) r^2-3L^2 (n-1) n \right]+\frac{ f' f''r^4}{8}n(n-1)\left( f r^2- L^2 \right)\Bigg]=\omega^3\, ,
\end{align}
where $\omega^3$ is an integration constant related to the total energy ---see below.  

Assuming a $1/r$ expansion of the function $f$ in the asymptotic limit, we find the following result
\begin{equation}\label{eq:V3dasympt}
V(r)=\chi_0\frac{r^2}{L^2}-1+\frac{\omega^3}{\Upsilon'(\chi_0) r L^2}+\mathcal{O}\left(\frac{1}{r^2}\right)\, ,
\end{equation}
where $\chi_0$ is a constant determined from \req{polyh} and $\Upsilon(\chi)$ is the ``characteristic polynomial" defined in \req{upside}.
From this asymptotic solution we can already identify the two integration constants $N$ and $\omega^3$. First, we see that the boundary metric at $r\rightarrow\infty$ reads
\begin{equation}
\diff s_{\rm bdry}^2\big|_{r\rightarrow\infty}=\frac{r^2\chi_0N^2}{L^2}\left[-\diff t^2+\frac{L^2}{\chi_0N^2} \diff \Xi^2\right]\, .
\end{equation}
Therefore, the $2+1$ boundary theory lives in the space $\mathbb{R}\times \mathbb{H}^2$, where the radius $R$ of the hyperbolic factor is $L/(\sqrt{\chi_0}N)$, so we have
$
N=L/ (\sqrt{\chi_0}R)
$. 
On the other hand, $\omega^3$ is related to the total energy of the spacetime as can be verified using the well-known extensions of the ADM formula to higher-order gravities \cite{Abbott:1981ff,Deser:2002jk}. Taking into account that the effective Newton's constant of the theory in this AdS background is
$
G_{\rm eff}=-G/\Upsilon'(\chi_0)
$,  we obtain $\omega^3=8\pi GL\sqrt{\chi_0}R E /( V_{\rm \mathbb{H}^2})$, where $V_{\rm \mathbb{H}^2}$ is the regularized volume of the unit hyperbolic space~\cite{CHM}.

Let us now analyze the behavior of $V(r)$ near the horizon. For that, we assume a series expansion of the form
\begin{equation}\label{nhexpa}
V(r)=\frac{4\pi T}{N}(r-\rh)+\sum_{n=2}^{\infty}a_n (r-\rh)^n\, ,
\end{equation}
near some undetermined point $r=\rh$. In this expansion we are already identifying $V'(\rh)$ with the temperature of the black hole, which is  defined as the inverse of the periodicity of the Euclidean time $\tau=i t$. When we insert this expansion into \req{eq:V3dEq}, we obtain an infinite number of equations for the coefficients of the series above. The first two equations are particularly relevant, since they only involve $\rh$, $T$ and $\omega^3$. None of the $a_n$ appear. They can be written in a convenient way as follows:
\begin{align}
\rh \left[\Upsilon'(\chi) \left(1-\frac{\rh^2}{L^2} \chi\right)+\frac{\rh^2}{L^2} \Upsilon(\chi)\right]=\omega^3\, ,\quad 
2 \Upsilon'(\chi) \left(L^2-\rh^2 \chi \right)+6 \rh^2 \Upsilon(\chi)=0\, ,
\end{align}
where we have introduced the notation
\begin{equation}
\chi\equiv \frac{2\pi T L^2}{N \rh}\, .
\end{equation}
These equations can be solved in order to obtain $\rh$, $T$ and $E$ in terms of $\chi$. We can write the answers fully in terms of $\Upsilon(\chi)$ as
\begin{align}
\label{eq:rhsol3d}
\rh=&L\left[\frac{\Upsilon'(\chi)}{\chi \Upsilon'(\chi)-3\Upsilon(\chi)}\right]^{1/2} \, ,\\
\label{eq:Esol3d}
E=&-\frac{V_{\rm \mathbb{H}^2}L^2}{4\pi G\sqrt{\chi_0}R} \left[\frac{\Upsilon'(\chi)}{\chi \Upsilon'(\chi)-3\Upsilon(\chi)}\right]^{3/2} \Upsilon(\chi)\, ,\\
\label{eq:Tsol3d}
T=&\frac{ \chi }{2\pi \sqrt{\chi_0}R }\left[\frac{\Upsilon'(\chi)}{\chi \Upsilon'(\chi)-3\Upsilon(\chi)}\right]^{1/2} \, .
\end{align}
Thus, by giving values to $\chi$ we parametrically generate the different relations $E(T)$, $T(\rh)$, and so on.\footnote{It is an interesting fact that all these quantities, as well as the entropy, can be written in terms of the characteristic polynomial $\Upsilon$ and its derivatives. A similar phenomenon occurs for Lovelock theories in general dimensions \cite{Camanho:2010ru,Camanho:2011rj,Paulos:2011zu,Camanho:2015ysa}, and presumably extends to more general QT and GQT theories. This will be subject of future study.}

Let us now compute the entropy of the solutions using Wald's formula \cite{Wald:1993nt,Iyer:1994ys}. This is given by
\begin{equation}\label{wald}
S=-2\pi\int_{\mathcal{H}}d^2x\sqrt{h}P_{abcd}\epsilon^{ab}\epsilon^{cd}\, ,\quad \text{where} \quad P_{abcd}\equiv\frac{\partial \mathcal{L}}{\partial R^{abcd}}\, ,
\end{equation}
and $\epsilon^{ab}$ is the binormal to the horizon,  normalized so that $\epsilon^{ab}\epsilon_{ab}=-2$.
For the metric \eqref{eq:hypBH3d1}, this formula can be simplified to yield
\begin{equation}
S=8\pi \rh^2 V_{\rm \mathbb{H}^2}P_{tr}^{\,\,\,\, tr}\Big|_{r=\rh}\, .
\end{equation}
We can evaluate this expression using the results in \cite{Bueno:2019ycr}, where an explicit expression for $P_{abcd}$ was provided. We find
\begin{equation}
P_{tr}^{\,\,\,\, tr}=\frac{1}{16\pi G}\left[\frac{1}{2}+\frac{1}{4}\sum_{n=3}^{\infty}\lambda_n n \frac{L^2}{r^2} \left(\frac{L^2V'}{2r}\right)^{-2+n} \left(2 (V+1) \frac{(n-1)}{n-2}- r V'\right)\right]\, .
\end{equation}
Evaluating this expression at $r=\rh$ and introducing the parameter $\chi$ we can write the entropy in terms of $\Upsilon(\chi)$ as 
\begin{equation}\label{eq:Ssol3d}
S=\frac{V_{\rm \mathbb{H}^2}L^2}{4 G}\left[\frac{\Upsilon'(\chi)^2}{3\Upsilon(\chi)-\chi \Upsilon'(\chi)}+\int_{0}^{\chi}dx\frac{\Upsilon''(x)}{x}\right] \, ,
\end{equation}
where we made use of \eqref{eq:rhsol3d}. It is now possible to check ---using \eqref{eq:Ssol3d}, \eqref{eq:Esol3d} and \eqref{eq:Tsol3d}--- that the first law of thermodynamics holds,
\begin{equation}
\diff E=T \diff S\, .
\end{equation}
The above expressions analytically capture the thermodynamic properties of an infinite family of higher-curvature hyperbolic black holes in a remarkably condensed fashion. This is a manifestation of the special properties of GQT gravities. 

With this information at hand, we are ready to evaluate $h_q$ from \req{sex}. 
%
%
Using the values of the temperature and the energy, given respectively by \eqref{eq:Tsol3d} and \eqref{eq:Esol3d}, we can write $h_q$ parametrically as
\begin{align}\label{hqqq}
h_q=\frac{L^2\Upsilon'(\chi)\Upsilon(\chi)}{4 G\chi(\chi \Upsilon'(\chi)-3\Upsilon(\chi))}\, ,\quad
q=\frac{\sqrt{\chi_0}}{ \chi }\left[\frac{\Upsilon'(\chi)}{\chi \Upsilon'(\chi)-3\Upsilon(\chi)}\right]^{-1/2} \, .
\end{align}
As a check, we see that for $\chi=\chi_0$ we get $q=1$ and $h_1=0$, as expected on general grounds. On the other hand, we observe that all derivatives of $h_q$ at $q=1$ are related to derivatives of $\Upsilon$ at $\chi_0$. In particular, the first and second derivatives read
\begin{align}
\left.\partial_q h_q \right|_{q=1}=-\frac{\Upsilon'(\chi_0)}{8\chi_0 G}\, ,\quad
\left.\partial^2_q h_q \right|_{q=1}=\frac{14 \Upsilon'(\chi_0)+7\chi_0 \Upsilon''(\chi_0)}{64\chi_0 G}\, .
\end{align}
Then, using relation (\ref{hq2t}), which connects $\left.\partial^2_q h_q \right|_{q=1}$ to $t_4$ for general CFTs, we finally obtain an expression for $t_4$ valid for the infinite class of GQT theories considered. This takes the simple form
\begin{equation}\label{eq:t43d}
t_4=210\chi_0\frac{\Upsilon''(\chi_0)}{\Upsilon'(\chi_0)}\, .
\end{equation}
It reduces to the one obtained for Einsteinian cubic gravity in \cite{Bueno:2018xqc}.
The analogous relation between $\left.\partial_q h_q \right|_{q=1}$ and $\ctt$ given in \req{hq1} is in turn compatible with the general expression for $\ctt$ given in \req{cte}, which 
in $d=3$ reads
$
\ctt=-3(L/\sqrt{\chi_0})^2 \Upsilon'(\chi_0)/ (\chi_0\pi^3 G) \, .
$

Now let us consider the holographic free energy for CFTs on squashed spheres. As we have seen, for all theories in \eqref{genGQTG3d} such free energy is given by \req{fee0e} evaluated for $d=3$, which reads
\begin{equation}
F_{\mathbb{S}^3_{\varepsilon}}=-\frac{4\pi^2 L^4 }{3} \frac{\mathcal{L}\left[\chi_0/(1+\varepsilon)\right]}{\left[\chi_0/(1+\varepsilon)\right]^2}\, .
\end{equation}
Expanding this expression around $\varepsilon=0$ and using the relation between the functions $\mathcal{L}(\chi)$ and $\Upsilon(\chi)$ in \eqref{upside}, as well as \req{polyh}, we obtain
\begin{equation}
F_{\mathbb{S}^3_{\varepsilon}}=F_{\mathbb{S}^3}+\frac{\pi L^2 \Upsilon'(\chi_0)}{2\chi_0 G}\varepsilon^2-\frac{\pi L^2 \Upsilon''(\chi_0)}{6 G}\varepsilon^3+\mathcal{O}(\varepsilon^4)\, .
\end{equation}
Then, using \eqref{eq:t43d} and \eqref{cte}, we can write this expansion in terms of $\ctt$ and $t_4$. The result reads
\begin{equation}
F_{\mathbb{S}^3_{\varepsilon}}=F_{\mathbb{S}^3}-\frac{\ctt \pi^4}{6}\varepsilon^2\left[1-\frac{t_4}{630}\varepsilon+\mathcal{O}(\varepsilon^2)\right]\, ,
\end{equation}
which is in perfect agreement with the conjectural relation in \req{3conj} proposed in \cite{Bueno:2018yzo}. As mentioned above, this was originally proposed using the Einsteinian cubic gravity result and then cross-checked against numerical results  \cite{Bobev:2017asb} corresponding to a free fermion and a conformally-coupled scalar. The fact that it holds for an infinite family of holographic higher-order gravities provides strong evidence in favor of its validity for general CFTs. 



Before closing the section, let us make an additional observation. In order to obtain the results above, we used, as an intermediate step, the connection between $t_4$ and the second derivative of the twist operators scaling dimension $h_q$. In fact, our computations show the existence of an equivalence between  $F_{\mathbb{S}^3_{\varepsilon}}$ and $h_q$ which holds at least for the class of theories considered here. This is made more explicit if we take the first derivative of  $F_{\mathbb{S}^3_{\varepsilon}}$, which reads
\begin{equation}\label{f1ups}
F^{(1)}_{\mathbb{S}^3_{\varepsilon}}=-\frac{\pi L^2}{G\chi_0}\frac{(1+\varepsilon)}{\chi_0}\Upsilon\left[\frac{\chi_0}{1+\varepsilon}\right]\, .
\end{equation}
Comparing with \req{hqqq}, we observe that both $F_{\mathbb{S}^3_{\varepsilon}}$ and $h_q$ are determined by the function $\Upsilon\left(\chi\right)$, and hence both contain the same information. Establishing a direct relation between the two quantities is complicated, but nevertheless we can derive simple relations between their derivatives. As we show explicitly in Appendix \ref{hFs}, these relate $F^{(j)}_{\mathbb{S}^3_{\varepsilon}}(0)$ to $h_q^{(j-1)}(1),\dots,h_q^{(1)}(1)$. It is very tempting to speculate with the possibility that those relations may extend to general CFTs. We test this using analytic results for the quantities involved in the case of a free fermion, and find that the predicted relation between $F^{(4)}_{\mathbb{S}^3_{\varepsilon}}(0)$ and $h_q^{(3)}(1),h_q^{(2)}(1)$ and $h_q^{(1)}(1)$ ---which appears in \req{Pablonian4}--- is not satisfied (the ones for $j=1,2,3$ do hold, in agreement with the rest of general results/conjectures of the paper).

\section{Five-dimensional holographic CFTs}\label{5dtheor}
In this section we use holographic calculations for quartic and quintic GQT theories as well as QT theories of arbitrary orders to establish a new relation between $F_{\mathbb{S}^5_{\varepsilon}}^{(3)}(0)$ and a linear combination of the stress-tensor three-point charges $t_2$ and $t_4$, which we conjecture to hold for general five-dimensional CFTs. 
\subsection{Quartic GQT theories}
Let us start our study of five-dimensional CFTs by analyzing the quartic theories for which explicit AdS-Taub-NUT solutions were constructed in \cite{NewTaub2}. The free energy of those solutions was also computed in the same paper, and it was later observed \cite{Bueno:2018yzo} that the resulting expressions match the general formula \eqref{fee0e}. The Euclidean action of the theory is given by
\begin{align}\label{full6}
I_E = -\int \frac{\diff^6 x \sqrt{g}}{16 \pi G} \left[ \frac{20}{L^2} + R + \frac{\lambda_{\rm \ssc GB} L^2 }{6} {\cal X}_4 + L^6 \left(\xi   \mathcal{S} + \zeta \mathcal{Z} \right)\right] \, ,
\end{align}
where we have included the usual Gauss-Bonnet density $\mathcal{X}_4\equiv R^2-4R_{ab}R^{ab}+R_{abcd}R^{abcd}$, and 
\begin{align}
{\cal S} =& \frac{-1}{216}\big[ 992 R_{a}{}^{c} R^{ab} R_{b}{}^{d} R_{cd} + 28 R_{ab} R^{ab} R_{cd} R^{cd} - 192 R_{a}{}^{c} R^{ab} R_{bc} R - 108 R_{ab} R^{ab} R^2 
\\
&+ 1008 R^{ab} R^{cd} R R_{acbd} + 36 R^2 R_{abcd} R^{abcd} - 2752 R_{a}{}^{c} R^{ab} R^{de} R_{bdce} + 336 R R_{a}{}^{e}{}_{c}{}^{f} R^{abcd} R_{bedf} 
\nonumber\\
&- 168 R R_{ab}{}^{ef} R^{abcd} R_{cdef} - 1920 R^{ab} R_{a}{}^{cde} R_{b}{}^{f}{}_{d}{}^{h} R_{cfeh} + 152 R_{ab} R^{ab} R_{cdef} R^{cdef} 
\nonumber\\
&+ 960 R^{ab} R_{a}{}^{cde} R_{bc}{}^{fh} R_{defh} - 1504 R^{ab} R_{a}{}^{c}{}_{b}{}^{d} R_{c}{}^{efh} R_{defh} + 352 R_{ab}{}^{ef} R^{abcd} R_{ce}{}^{hi} R_{dfhi} 
\nonumber\\
&- 2384 R_{a}{}^{e}{}_{c}{}^{f} R^{abcd} R_{b}{}^{h}{}_{e}{}^{i} R_{dhfi} + 4336 R_{ab}{}^{ef} R^{abcd} R_{c}{}^{h}{}_{e}{}^{i} R_{dhfi} - 143 R_{ab}{}^{ef} R^{abcd} R_{cd}{}^{hi} R_{efhi} 
\nonumber\\
&- 436 R_{abc}{}^{e} R^{abcd} R_{d}{}^{fhi} R_{efhi} + 2216 R_{a}{}^{e}{}_{c}{}^{f} R^{abcd} R_{b}{}^{h}{}_{d}{}^{i} R_{ehfi} - 56 R_{abcd} R^{abcd} R_{efhi} R^{efhi} \big] \, ,
\nonumber \\ 
{\cal Z} =&\frac{-1}{144}\big[  -112 R_{a}{}^{c} R^{ab} R_{b}{}^{d} R_{cd} - 36 R_{ab} R^{ab} R_{cd} R^{cd} + 18 R_{ab} R^{ab} R^2 - 144 R^{ab} R^{cd} R R_{acbd} 
\\
&- 9 R^2 R_{abcd} R^{abcd} + 72 R^{ab} R R_{a}{}^{cde} R_{bcde} + 576 R_{a}{}^{c} R^{ab} R^{de} R_{bdce} - 400 R^{ab} R^{cd} R_{ac}{}^{ef} R_{bdef} 
\nonumber\\
&+ 48 R R_{a}{}^{e}{}_{c}{}^{f} R^{abcd} R_{bedf} + 160 R_{a}{}^{c} R^{ab} R_{b}{}^{def} R_{cdef} - 992 R^{ab} R_{a}{}^{cde} R_{b}{}^{f}{}_{d}{}^{h} R_{cfeh} 
\nonumber\\
&+ 18 R_{ab} R^{ab} R_{cdef} R^{cdef} - 8 R^{ab} R_{a}{}^{cde} R_{bc}{}^{fh} R_{defh} + 238 R_{ab}{}^{ef} R^{abcd} R_{ce}{}^{hi} R_{dfhi} 
\nonumber\\
&- 376 R_{a}{}^{e}{}_{c}{}^{f} R^{abcd} R_{b}{}^{h}{}_{e}{}^{i} R_{dhfi} + 1792 R_{ab}{}^{ef} R^{abcd} R_{c}{}^{h}{}_{e}{}^{i} R_{dhfi} - 4 R_{ab}{}^{ef} R^{abcd} R_{cd}{}^{hi} R_{efhi} 
\nonumber\\
&- 284 R_{abc}{}^{e} R^{abcd} R_{d}{}^{fhi} R_{efhi} + 320 R_{a}{}^{e}{}_{c}{}^{f} R^{abcd} R_{b}{}^{h}{}_{d}{}^{i} R_{ehfi} \big]\, , \nonumber
\end{align}
are two canonically-normalized quartic GQT densities \cite{Ahmed:2017jod}. In particular, ${\cal Z}$ belongs to the QT subfamily, as it modifies the equation of $f(r)$ for static black holes algebraically. 
On the other hand, ${ \cal S}$ contributes to such equation with up to two derivatives of $f(r)$, so it is a standard GQT density.  As we have mentioned, the free energy of the CFT$_5$ dual to \eqref{full6} on $\mathbb{S}^5$ was computed in \cite{NewTaub2,Bueno:2018yzo}. The result reads
\begin{equation}\label{eqfequarticd=5}
F_{\mathbb{S}_\varepsilon^{5}} = \frac{\pi^2 L^4 (1+\varepsilon)^3}{ G \chi_0^3 } \left[\frac{2}{3} - \frac{\chi_0}{1+\varepsilon} + \frac{2 \lambda_{\ssc \rm GB} \chi_0^2}{(1+\varepsilon)^2} - \frac{2 (\xi + \zeta) \chi_0^4}{(1+\varepsilon)^4} \right] \, ,
\end{equation}
in agreement with \req{LL}.

In order to identify a possible generalization of \req{3conj} valid for $d=5$ CFTs, we should expand \req{eqfequarticd=5} around $\varepsilon=0$ and express the third derivative,
\begin{equation}\label{F35dholo}
F^{(3)}_{\mathbb{S}_\varepsilon^{5}}(0)=\frac{4 \pi ^2 L^4 \left(1+3 \chi_0^4 (\xi+\zeta)\right)}{G \chi_0^3}\, ,
\end{equation}
in terms of a linear combination of $\ctt$, $\ctt t_2$ and $\ctt t_4$. 
The two-point function charge $\ctt$ is given by the general formula  \eqref{cte}, and therefore reads
\begin{equation}
\ctt= \left[1-2\chi_0\lambda_{\rm \ssc GB}- 4\chi_0^3(\chi+\zeta) \right]\frac{30 (L/\sqrt{\chi_0})^4}{\pi^4G}\, .
\end{equation}
On the other hand, in order to compute $t_2$ and $t_4$, we use the holographic energy-flux calculation described in Section \ref{3pff}. Evaluating the action (\ref{full6}) on-shell for the perturbed metric (\ref{eq:perturbationsAdSD}), we obtain, after some massaging,
\begin{equation}\label{eq:onshellquartic}
\begin{aligned}
I_E=-\frac{1}{16\pi G}\int d^5y du \frac{{\tilde L}^{4}W\phi \partial_-^2\phi}{u^4} \Big[&1-2\lambda_{\ssc \rm GB}\chi_0-4\left(\xi+\zeta\right)\chi_0^3\\
&+\left(\frac{\lambda_{\ssc \rm GB}\chi_0}{3}+\left(2\xi-182\zeta\right) \chi_0^3\right) T_2 - 36\zeta \chi_0^3  T_4 \Big]\ ,
\end{aligned}
\end{equation}
where 
\begin{align}
T_2&\equiv \frac{u^2\left(\partial_1^2 W+\partial_2^2 W\right)-2u \,\partial_u W}{W}\, , \\
T_4&\equiv u^2\frac{ \partial_3^2W+7  \partial_u W/u-4\left( \partial_1^2W+\partial_2^2W\right)}{W}+u^3\frac{\partial_u\partial_1^2W+\partial_u\partial_2^2W-u\partial_1^2\partial_2^2 W}{W}\, .
\end{align}
Evaluating $T_2$ and $ T_4$ for $W$ given by eq.~ (\ref{eq:solW}) 
we get
\begin{equation}
\label{eq:T26Devaluated}
T_2=40\left[\frac{n_1^2+n_2^2}{2}-\frac{1}{4}\right]\, , \quad
T_4=-420\left[2n_1^2n_2^2-\frac{1}{12}\right]\, .
\end{equation}
Plugging this result in the action and comparing with \req{eq:flux}, we read off the flux parameters. The result is
\begin{equation}\label{t2d=5}
t_2= \frac{40}{3}\frac{\lambda_{\rm \ssc GB}\chi_0+\left(6\xi-546\zeta\right)\chi_0^3}{\left[1-2\lambda_{\rm \ssc GB}\chi_0-4 \left(\xi+\zeta\right)\chi_0^3\right]}\, , \quad
t_4=\frac{15120\,\zeta \chi_0^3}{\left[1-2\lambda_{\rm \ssc GB}\chi_0-4 \left(\xi+\zeta\right)\chi_0^3\right]}\, .
\end{equation}
This reduces to the Gauss-Bonnet result for $\zeta=\xi=0$ \cite{Buchel:2009sk}.

As a consistency check, we have considered the hyperbolic black holes of the theory and obtained the following expression for the twist-operator scaling dimension near $q=1$, 
\begin{align}\label{eqhqd=5conS}
h_q=&+\frac{ \left[ 1- 2 \chi_0 \lambda_{\rm \ssc GB}-4 \chi_0^3 (\xi + \zeta) \right] (L/\sqrt{\chi_0})^4}{16
	 G}(q-1) \\\nonumber 
	&-\frac{  \left[31-26\chi_0 \lambda_{\rm \ssc GB} +92\chi_0^3(\xi+\zeta)\right] (L/\sqrt{\chi_0})^4}{512  G }(q-1)^2 + \mathcal{O}(q-1)^3\, .
\end{align}
Comparing with \req{hq1} and \req{hq2t}, we find that the values of $t_2$ and $t_4$ obtained in \req{t2d=5} agree with this expression.

Having computed $t_2$ and $t_4$, we are ready to write the desired expansion for $F_{\mathbb{S}_\varepsilon^{5}}$. One can check that, indeed, it is  possible to express \eqref{F35dholo} as a combination of $\ctt$, $\ctt t_2$ and $\ctt t_4$.\footnote{To show this one needs to take into account the embedding equation satisfied by $\chi_0$: $1-\chi_0+\lambda_{\rm \ssc GB}\chi_0^2+(\xi+\zeta)\chi_0^4=0$.} This is a nontrivial fact which was not guaranteed a priori.  Thus, the expansion of $F_{\mathbb{S}_\varepsilon^{5}}$ up to cubic order in $\varepsilon$ reads
\begin{equation}\label{5conj2}
F_{\mathbb{S}_\varepsilon^5}=F_{\mathbb{S}_0^5} + \frac{\ctt \pi^6}{30}\varepsilon^2 + \frac{\ctt \pi^6}{45}\left[1 + \frac{3}{40} t_2 + \frac{23}{630}t_4\right]\varepsilon^3+ O(\varepsilon^4)\, .
\end{equation}
Naturally, the leading correction to the round-sphere result agrees with the general-CFT one appearing in \req{f20}. On the other hand, the  subleading piece is a new prediction which we conjecture to be valid for general theories. The rest of the section will be devoted to testing this conjecture. We observe that while the constant piece differs, the relative coefficients between the $t_2$ and $t_4$ terms precisely agree with the ones appearing in $\left.\partial^2_q h_q\right|^{d=5}_{q=1} $ ---see \req{hq2eval}. This intriguing coincidence implies that we can test \req{5conj2} for additional higher-curvature theories without computing $t_2$ and $t_4$ separately. We can instead evaluate $h_q$, identify the linear combination $3 t_2 /40+23 t_4/630 $, and then verify whether or not $F_{\mathbb{S}_\varepsilon^5}$ satisfies \req{5conj2} for the corresponding theory.
 

%


\subsection{Quintic GQT theories}

Unfortunately, we do not have at our disposal a complete understanding of all the possible GQT theories admitting Taub-NUT solutions in $d = 5$. Moreover, as we are going to see, for $d > 3$ there exist multiple \textit{distinct} GQT theories at a given order $n > 3$.\footnote{This is a feature which had been previously overlooked in the literature.} 
In order to provide additional evidence for 
 \req{5conj2}, here we consider quintic gravities admitting Taub-NUT solutions. 

Our approach for constructing the quintic theories is in line with previous methods outlined in, \eg~\cite{PabloPablo4, Arciniega:2018tnn}. We begin with a basis of invariants including terms up to quintic order in curvature ---see Appendix~\ref{moreQuintic}. We construct from these invariants the most general combination that is quintic in curvature and then constrain the couplings so that the theory admits single-function Taub-NUT solutions of the form (\ref{FFnut}). This amounts to imposing  $\delta I/\delta V=0$
on the action. After imposing this condition, we evaluate the same Lagrangian density on a static, spherically symmetric background. This allows the theories to be classified as either QT or GQT. We find that the theories decouple into three objects: a QT gravity, and two \textit{distinct} GQTGs (in the sense that the field equations following from these densities are independent). We restrict our attention now to the GQT theories ---a general treatment of the QT case will appear in the next subsection. Including only the Einstein-Hilbert piece along with the quintic GQTG terms, the action reads 
\be 
I_E=-  \frac{1}{16 \pi G} \int \diff^{6}x \sqrt{|g|} \left[\frac{20}{L^2} + R +   L^8 \left(\beta \mathcal{Q}_1 + \mu \mathcal{Q}_2 \right) \right] \, ,
\ee
where the canonically-normalized densities $\mathcal{Q}_1$ and $\mathcal{Q}_2$ are presented in Appendix~\ref{moreQuintic}. 

We have computed the field equations for this theory evaluated on the Taub-NUT ansatz, however, for our purposes here it will not be necessary to perform an analysis of the solutions of these field equations to the level of detail presented in~\cite{NewTaub2}. Furthermore, due to the sheer complexity of the resulting expressions we do not present them here. 
It is possible to compute the free energy of Taub-NUT solutions in these theories using the techniques of~\cite{HoloECG} with the modifications described in~\cite{NewTaub2}. For this we need only know that, in the vicinity of the NUT, the behavior of the metric function is
\be 
V_{\mathbb{CP}^2}(r) = \frac{r-n}{3 n} + \mathcal{O}(r-n)^2  \, .
\ee
A simple, if somewhat tedious, computation making use of the Euclidean on-shell action 
reveals that the conjectural formula for the free energy holds also for the quintic theories included here, that is 
\be 
F_{\mathbb{S}_\varepsilon^5} = \frac{\pi^2 L^4 (1 + \varepsilon)^3}{G \chi_0^3} \left[\frac{2}{3} - \frac{\chi_0}{1+\varepsilon} - \frac{(\beta + \mu) \chi_0^5}{(1+\varepsilon)^5} \right] \, .
\ee
From this we can easily extract the third derivative:
\be 
F_{\mathbb{S}_\varepsilon^5}^{(3)}(0) = \frac{4 \pi^2 L^4 (1 + 5(\mu + \beta) \chi_0^4}{\chi_0^2 G} = -\frac{4 \pi^2 (L/\sqrt{\chi_0})^4}{G} \Upsilon'(\chi_0) \left[1- \frac{\chi_0}{2} \frac{\Upsilon''(\chi_0)}{\Upsilon'(\chi_0)} \right] \, ,
\ee
where in the second line we wrote the result in terms of the embedding function, $\Upsilon(\chi)=1-\chi+(\beta+\mu)\chi^5$.

In order to test the validity of \req{5conj2},
 we must have at hand the flux parameters for these theories.  We will access the relevant linear combination from the second derivative of 
 $h_q$, as described above. For this, we need an understanding of the thermodynamics of hyperbolic black holes. These are described by the following metric,
 \begin{equation}\label{eq:hypBHquintic}
 \diff s^2=-N(r)^2V(r)\diff t^2+\frac{\diff r^2}{V(r)}+r^2 \diff \Xi_{(4)}^2\, ,  \quad V(r)\equiv \frac{r^2}{L^2}f(r)+k\, ,
 \end{equation}
where now $\diff \Xi_{(4)}^2$ is the metric of the unit four-dimensional hyperbolic space $\mathbb{H}^4$.
The field equations are reduced to a single equation for $f$ which reads
\be 
\frac{\omega^5}{r^5} =1-f(r)+ \frac{L^2}{4r^5}\left[ \mu \mathcal{F}_{\mathcal{Q}_1} + \beta \mathcal{F}_{\mathcal{Q}_2}\right]  \, ,
\ee
where the quintic contributions to the field equations are presented in the appendix and $\omega^5$ is proportional to the ADM energy of the solution. On the other hand, $N=$ constant, as usual. 

From these equations, the computation of $h_q$ proceeds in exactly the same fashion as in the previous sections. For brevity, here we note only that the intermediate result for $x_q$ reads
\be 
x_q = 1 - \frac{(q-1)}{4} - \frac{5}{128} \frac{(12 \chi_0 -5)(q-1)^2}{4 \chi_0- 5} + \mathcal{O}(q-1)^3 \, ,
\ee
which is used in arriving at the final result for the second derivative of $h_q$. We find
\be 
\left.\partial_q^2 h_q\right|_{q=1}
= \frac{L^4 \left[31 \Upsilon'(\chi_0) - 18\chi_0 \Upsilon''(\chi_0) \right]}{256 \chi_0^2 G}\, ,
\ee
where again we wrote the result in terms of derivatives of the characteristic function $\Upsilon$. Using Eq.~\eqref{hq2eval} we then solve for the combination
\be 
\frac{3}{40}t_2 + \frac{23}{630}t_4 = - \frac{\chi_0}{2} \frac{\Upsilon''(\chi_0)}{\Upsilon'(\chi_0)} \, . 
\ee
Using this in the result for the free energy we can see that the quintic theories predict
\be  \label{5conj3}
F_{\mathbb{S}_\varepsilon^5}^{(3)}(0) = \frac{2 \pi^6 C_T}{15}  \left[1 +  \frac{3}{40}t_2 + \frac{23}{630}t_4\right] \, ,
\ee
in precise agreement with the result obtained for the quartic ones in the previous subsection.

\subsection{General QT theories}
As we have emphasized before, the equations of motion of QT theories evaluated on static black hole solutions with various horizon topologies are algebraic for the metric function, and particularly simple \cite{Quasi2,Quasi,Dehghani:2011vu,Ahmed:2017jod,Cisterna:2017umf}. In particular, consider a general $(d+1)$-dimensional QT action of the form 
\begin{equation}\label{actionQT}
I_E= -\int \frac{\diff^{d+1} x \sqrt{g}}{16 \pi G} \left[ \frac{d(d-1)}{L^2} + R +\sum_n \lambda_n L^{2(n-1)} \mathcal{Z}_{(n)} \right]\, ,
\end{equation}
where the $\mathcal{Z}_{(n)}$ are order-$n$ QT densities. These were explicitly shown to exist for arbitrary $n$ in \cite{Bueno:2019ycr} ---see that paper for an explicit formula for $ \mathcal{Z}_{(n)}$. 

 The equations of motion for an ansatz of the form
\begin{equation}\label{eq:hypBH3d}
\diff s^2=-N(r)^2V(r)\diff t^2+\frac{\diff r^2}{V(r)}+r^2 \diff \Sigma_{(k)}^2\, ,  \quad V(r)\equiv \frac{r^2}{L^2}f(r)+k\, ,
\end{equation}
where $ \diff \Sigma_{(k)}^2$ denotes the arc element of a unit $(d-1)$-dimensional sphere/hyperbolic plane/Euclidean space for $k=1,-1,0$ respectively, reduce to  $N=$ constant, and
\be 
\frac{\omega^{d}}{r^{d}} =1-f(r)+\sum_n  \lambda_n f(r)^n\, ,
\ee
where the integration constant is related to the ADM energy of the solution \cite{Arnowitt:1960zzc,Arnowitt:1960es,Arnowitt:1961zz,Deser:2002jk} as 
\be
E=\frac{(d-1)\omega^{d} N V_{\Sigma}}{16 \pi G L^2}\, .
\ee
In terms of the characteristic function defined in \req{upside}, the above equation takes an even simpler form, namely
\be 
\frac{\omega^{d}}{ r^{d}} =  \Upsilon(\chi) \, , \quad \text{where} \quad \chi \equiv f(r)\, .
\ee
Considering a near-horizon expansion of $f(r)$ as in \req{nhexpa}, we obtain two equations for $\omega$ and $T$, which read
\begin{align}\label{omT}
\omega^{d}= \rh^{d}  \Upsilon(\chi_{\rm h})\,, \quad 
T= -\frac{N}{2\pi} \left[\frac{k}{\rh}+\frac{d\,\rh \Upsilon(\chi_{\rm h})}{2L^2  \Upsilon'(\chi_{\rm h})} \right]\, ,
\end{align}
where here $\chi_{\rm h} \equiv f(\rh) = -L^2 k/\rh^2$. 
These expressions properly reduce to the previously known ones corresponding to: $n=3$ for general $d$ \cite{Quasi2,Quasi}, $n=4$ for general $d$ \cite{Dehghani:2011vu,Ahmed:2017jod}, and $n=5$ in $d=4$ \cite{Cisterna:2017umf}.

Let us now see what happens with the entropy. For this, we use Wald's formula \req{wald}. 
%
As before, we only need $P_{tr}{}^{tr}$, which can be obtained from the general expression for $P_{ab}{}^{cd}$ computed in \cite{Bueno:2019ycr}.
We find
\be 
P_{tr}{}^{tr}  = - \frac{1}{2} \frac{\partial {\cal L}|_V}{\partial V''} \, , \quad \text{so} \quad S = - 4 \pi V_{\Sigma} \rh^{d-1} \left[ \frac{\partial {\cal L}|_V}{\partial V''} \right]_{r=\rh}\, ,
\ee
where for QT gravities one finds 
\be 
\frac{\partial {\cal L}|_V}{\partial V''} = \frac{1}{r^{d-1}} \left[\frac{L^2 \rh^{d-1}}{d(d+1)}  {\cal L}'(\chi_{\rm h}) \right]  \quad \Rightarrow \quad S =  \frac{V_{\Sigma} \rh^{d-1}}{4G } \left[-\frac{16\pi L^2}{d(d+1)} {\cal L}'(\chi_{\rm h}) \right] \, .
\ee
This reproduces all the particular cases previously studied in \cite{Quasi2,Quasi,Dehghani:2011vu,Ahmed:2017jod,Cisterna:2017umf}. Using the above expressions it is possible to verify that the first law is satisfied, as it should.
Naturally, all expressions can be straightforwardly written  in terms of the $\lambda_n$ for a general QT Lagrangian  (\ref{actionQT}) using \req{hh} as usual. 

The above expressions for $E$, $T$ and $S$ are valid for general-order QT theories in arbitrary dimensions. Let us now go back to our original  motivations  ---namely, obtaining $h_q$ for five-dimensional CFTs dual to QT theories--- and therefore set $k=-1$ and $N=L/(\sqrt{\chi_0}R)$. We will keep $d$ general and set $d=5$ at the end.

The equation which determines $x_q$, $T(x_q)=T_0/q$, can be obtained easily from \req{omT}. The result is
\be \label{xqQT}
0 = \frac{2 \chi_0}{d \,x_q^2 } \left[\frac{x_q}{q} - 1 \right] \Upsilon'\left[\frac{\chi_0}{x_q^2} \right] +   \,\Upsilon \left[\frac{\chi_0}{x_q^2} \right]  \, .
\ee
In each case, one should select the real root which reduces to the Einstein gravity result when all higher-order couplings are set to zero. On the other hand, the expression for $h_q$ can be obtained from \req{sex}. We find 
\begin{align} \label{hqQT}
h_q &= - \frac{ qx_q^{d} (L/\sqrt{\chi_0})^{d-1}  }{8G \fin  } \Upsilon\left[\frac{\fin}{x_q^2} \right] \, .
\end{align}
Expanding around $q=1$, one finds for $x_q$
\be \label{xqexp}
x_q = 1 - \frac{(q-1)}{(d-1)}  + \frac{d\left[(2d-3) \Upsilon'(\fin) - 2 \fin \Upsilon''(\fin) \right]}{2(d-1)^3 \Upsilon'(\fin)} (q-1)^2 + {\cal O} \left(q-1 \right)^3 \, .
\ee
Using this and \req{hqQT}, we can also obtain an explicit expression for $h_q$ 
perturbatively around $q=1$. The result for the first nonvanishing terms reads
\begin{align}\label{hp1}
\left.\partial_q h_q \right|_{q=1}&=\frac{(L/\sqrt{\chi_0})^{d-1}}{4G(d-1)} \left[-\Upsilon'(\fin)\right]\, , \\ 
\left.\partial^2_q h_q \right|_{q=1}&=\frac{(L/\sqrt{\chi_0})^{d-1}}{4G(d-1)^3} \left[(2d^2 - 4d + 1) \Upsilon'(\fin) - 2(2d-1) \fin \Upsilon''(\fin) \right]\, .
\end{align}
Comparing with \req{hq1} and \req{cte} we observe that $\left.\partial_q h_q \right|_{q=1}$ agrees with the general formulas. On the other hand, comparing the result for $\left.\partial^2_q h_q \right|_{q=1}$ with \req{hq2t}, we obtain the following result
\begin{align}\label{astarr}
 t_2 + \frac{(7d^3 - 19 d^2 - 8 d + 8) }{(d-3)(d+1)(d+2)(2d-1)} t_4 &=- \frac{2d(d-1)\fin}{(d-2)(d-3)}  \frac{\Upsilon''(\fin)}{\Upsilon'(\fin)}  \, ,
\end{align}
which particularized to $d=5$ becomes again
\begin{align}\label{astarr}
\frac{3}{40} t_2 + \frac{23}{630} t_4 &=-\frac{\chi_0}{2} \frac{\Upsilon''(\fin)}{\Upsilon'(\fin)}  \, .
\end{align}
Now, expanding the general holographic formula \req{fee0e} around $\varepsilon=0$ we find
\begin{equation}
F_{\mathbb{S}_{\varepsilon}^5}^{(3)}(0)=\frac{-4\pi^2 (L/\sqrt{\chi_0})^4}{G} \Upsilon'(\chi_0) \left[1 -\frac{\chi_0}{2} \frac{\Upsilon''(\fin)}{\Upsilon'(\fin)}   \right]\, .
\end{equation}
This is expected to hold for all GQT theories admitting single-function AdS-Taub-NUT solutions, including the QT theories considered here. Therefore, using  \req{astarr} and \req{cte}  we are finally left with \req{5conj2}, \ie we find perfect agreement with our conjectural relation.


\section{Free fields}\label{freefieee}
The three-dimensional conjectural relation \req{3conj} was shown to hold numerically both for the scalar and the fermion in \cite{Bueno:2018yzo}. Here we confirm this expectation by computing $F_{\mathbb{S}^3_\varepsilon}^{(3)}(0)$ analytically in both cases. We also verify that the five-dimensional conformally-coupled scalar exactly verifies our new conjectural relation in \req{5conj2}, providing strong evidence for its validity. In addition, we explain how higher-(and lower-)order derivatives $F_{\mathbb{S}^d_\varepsilon}^{(k)}(0)$ can be obtained analytically in all cases.  We use a combination of heat-kernel and zeta-function regularization methods to obtain our results. 

The starting point is the free energy of conformally-coupled scalars and free Dirac fermions on an arbitrary Euclidean background and in general dimensions. This can be written as
\begin{equation}
F^{s}=(-1)^{2s}\frac{1}{2^{(1-2s)}}\log \det 
\mathfrak{D}_{s}
\, ,
\end{equation}
where $s=0$ for scalars and $s=1/2$ for fermions and where 
\begin{equation}
\mathfrak{D}_0\equiv -\nabla^2+ \frac{(d-2)}{4(d-1)}R\, , \quad \quad
\mathfrak{D}_{1/2}\equiv  i \slashed{D}\, ,
\end{equation}
are the conformal Laplacian and the Dirac operator respectively. This expression follows from the corresponding partition functions,
\begin{align}\label{Zscalar3d}
Z_{\rm 0}=\int \mathcal{D} \phi\, e^{-\frac{1}{2}\int d^dx \sqrt{g} \left[ (\partial \phi)^2 +\frac{ (d-2)}{4(d-1)}R\phi^2 \right]} \, , \quad Z_{\rm 1/2}=\int \mathcal{D} \psi\,  e^{-\int d^dx \sqrt{g} \left[\psi^{\dagger} (i \slashed{D} )\psi \right]}\, .
\end{align}
In $F^s$ there is an implicit scale, which can be made manifest by introducing a UV cutoff, $\mathfrak{D}_{s}\rightarrow \mathfrak{D}_{s}/\Lambda^{2(1-s)} $. Let us denote by $\lambda^{(s)}_{n,q_1,q_2,\dots}$ and $m_{n,q_1,q_2,\dots}^{(s)}$ the eigenvalues and corresponding multiplicities of the operator $\mathfrak{D}_{s}$. For general manifolds, those will involve several ``quantum numbers'', $n,q_1,q_2,\dots$, which for the rest of the subsection and the following we will collectively denote simply by $i$.  If we know the eigenvalues and their multiplicities, 
 $F^s$ can be formally written as
\begin{equation}\label{fseig}
F^s=(-1)^{2s}\frac{1}{2^{(1-2s)}} \sum_i m_i  \log \lambda_i\, ,
\end{equation}
where the sum over $i$ schematically represents sums over all indices and where the UV cutoff appears hidden but can be easily reintroduced by $\lambda_i \rightarrow \lambda_i/\Lambda^{2(1-s)}$. 
The above expression is divergent in general, and therefore requires regularization. 

\subsection{Heat-kernel and zeta-function regularizations}

Besides \req{fseig}, we can represent $F^s$ in additional ways, which can be useful for different purposes. For instance, we can define a ``heat-kernel'' 
\be  \label{hk}
K(t) \equiv \sum_i m_i e^{-t\lambda_i^{(2s+1)} } \, ,
\ee
and then write ---see \eg \cite{Vassilevich:2003xt,Bobev:2017asb,Anninos:2012ft,Bobev:2016sap},
\begin{equation} \label{fsheat}
F^s\rightarrow \frac{(-1)^{(2s+1)}}{2^{(1-2s)}(2s+1)}  \int_{\Lambda^{-2}}^{\infty} \frac{dt}{t} K(t)=\frac{(-1)^{(2s+1)}}{2^{(1-2s)}(2s+1)} \sum_i m_i \int_{\Lambda^{-2}}^{\infty} \frac{dt}{t} e^{-t \lambda_i^{(2s+1)} }\, ,
\end{equation}
which behaves as
\begin{equation}
\int_{\Lambda^{-2}}^{\infty} \frac{dt}{t} e^{-t \lambda_i^{(2s+1)} } =-\gamma -(2s+1) \log \left[\lambda_i/ \Lambda^{2(1-s)} \right] +\mathcal{O}(\lambda_i^{(2s+1)}/\Lambda^2)\, ,
\end{equation}
for $\lambda_{i} \ll \Lambda^{2(1-s)}$, where $\gamma $ is the Euler-Mascheroni constant, whereas
\begin{equation}
\int_{\Lambda^{-2}}^{\infty} \frac{dt}{t} e^{-t \lambda_i^{(2s+1)} } =e^{- \lambda_i^{(2s+1)}/\Lambda^2 } \left[\Lambda^2/\lambda^{(2s+1)}+\mathcal{O}(\Lambda^4/\lambda^{2(2s+1)}) \right] \, ,
\end{equation}
for $\lambda_{i} \gg \Lambda^{2(1-s)}$. Therefore, \req{fsheat} computes the required sum for modes smaller than $\Lambda$, while cutting it off exponentially above it. 

We can also define a ``spectral zeta function''~\cite{Hawking:1976ja} associated to $\mathfrak{D}_{s}$
\be \label{spectralzeta}
\myZ(p) \equiv \sum_i \frac{m_i}{\lambda_i^p}  \, ,
\ee
which converges for 
sufficiently large $p$. Formally we have
\be 
 \left. \myZ'(p) \right|_{p=0} = - \sum_i m_i \log \lambda_i\, , 
\ee
which, comparing with \req{fseig} makes the connection with $F^s$ obvious.
The objective, then, is to perform an analytic continuation of the spectral zeta function to obtain an expression valid at $p = 0$ so that the derivative can be computed. The analytic continuation is facilitated by the heat kernel defined in \req{hk},
which is related to the spectral zeta function via a Mellin transform 
\be 
G(p) \equiv \myZ(p) \Gamma(p) = \int_0^\infty t^{p-1} K(t) dt \, .
\ee
From the left-hand side of this expression, noting the asymptotic form of the gamma function, we have
\be \label{gsss}
G(p) = \frac{\myZ(0)}{p} - \gamma \myZ(0) + \myZ'(0) + \mathcal{O}(p) \, ,
\ee
and so we are able to extract the values of $\myZ(0)$ and $\myZ'(0)$.  On general grounds we know that the heat kernel will possess divergent pieces in the $t\to 0$ ($\Lambda \to \infty$) limit of the form
\be 
K(t) = \sum_{k=0}^{(d+1)/2} a_{d/2 - k} t^{-d/2 + k} + \mathcal{O}(t) \, ,
\ee
and, as a result, the integral for $G(p)$ is only well-behaved for $p > d/2$. To obtain an analytical continuation valid at $p = 0$ we follow ~\cite{Hawking:1976ja, Monin:2016bwf}, dividing the integration domain into intervals $[0, 1]$, $[1, \infty)$ while adding and subtracting the divergent parts. The final result is
\be \label{gpp}
G(p) = \int_0^1 dt \left[K(t) - \sum_{k=0}^{(d+1)/2} a_{d/2 - k} t^{-d/2 + k} \right] t^{p-1} + \sum_{k=0}^{(d+1)/2} \frac{a_{d/2 -k}}{p- d/2 + k} + \int_1^\infty dt K(t) t^{p-1} \, .
\ee
We note that $G(p)$ has no pole at $p = 0$ and therefore conclude that $\myZ(0) = 0$. Then, the $p = 0$ value of $G(p)$ will yield directly $\myZ'(0)$. In all the cases we consider below, the entire contribution arises from the $0$-limit of the first integral with all the additional terms cancelling amongst themselves.  


\subsection{Three-dimensional CFTs}
Let us now particularize the discussion to three-dimensional squashed spheres. The eigenvalues of the conformal Laplacian on $\mathbb{S}_{\varepsilon}^3$ are given by
\cite{Dowker:1998pi,Anninos:2012ft}
\begin{align}
\lambda^{(0)}_{n,q}(\varepsilon)&= n(n+2)-\frac{\varepsilon (n-2q)^2}{(1+\varepsilon)}  + \frac{(3-\varepsilon)}{4} \, , 
\end{align}
with $n=0,2,\ldots $ and  $q=0,1,\ldots,n$. The degeneracies are $m_{n,q}^{(0)}=n+1$. 

For the fermion, one has in turn \cite{Dowker:1998pi,Gibbons198098, HITCHIN19741}
\begin{align}\label{fermioneigen}
\lambda^{(1/2)}_{n,q,\pm}(\varepsilon) =\sqrt{1+\varepsilon}  \pm \frac{2\sqrt{n^2+4\varepsilon q (n-q)}}{\sqrt{1+\varepsilon}}  \, ,
\end{align}
where  $n$ and $q$ are integers. For the positive branch, denoted by ``$+$'', $n$ takes values from 1 to $\infty$ and $q$ from 0 to $n$, whereas for the negative branch, denoted by ``$-$'', $n$ goes from $2$ to $\infty$ and $q$  from $1$ to $n-1$. The degeneracies are $m_{n,q,\pm}^{(1/2)}=n$ for both branches. Sum over both of them must be performed when evaluating $ F_{\mathbb{S}_{\varepsilon}^3}^{1/2}$.  
Hence, the corresponding free energies can be written as
\begin{align}\label{feefe}
F^{0}_{\mathbb{S}^3_{\varepsilon}}&= +\frac{1}{2}\sum_{n=0}^{\infty} \sum_{q=0}^{n} (n+1) \log \lambda^{(0)}_{n,q}(\varepsilon) \, , \\ \label{feefefi} F^{1/2}_{\mathbb{S}^3_{\varepsilon}}&=-\left[ \sum_{n=1}^{\infty} \sum_{q=0}^{n} n \log \lambda^{(1/2)}_{n,q,+}(\varepsilon) +\sum_{n=2}^{\infty} \sum_{q=1}^{n-1} n \log \lambda^{(1/2)}_{n,q,-}(\varepsilon)  \right]\, ,
\end{align}
and analogous expressions can be written for the heat-kernel in \req{fsheat}.
While obtaining analytic expressions for the regularized parts of $F^{0}_{\mathbb{S}^3_{\varepsilon}}$ and $F^{1/2}_{\mathbb{S}^3_{\varepsilon}}$ seems to be a very difficult problem, our goal here will be to compute analytically the values of the coefficients  appearing in the respective expansions around $\varepsilon=0$.  In particular, this will allow us to perform exact tests of our conjectural relation \req{3conj}. We will consider different methods.


1. One possibility is to use expressions (\ref{feefe}) and (\ref{feefefi}) or, alternatively, their heat-kernel versions, ignore the infinite sums, take derivatives of the general term, evaluate them at $\varepsilon=0$, and only then, deal with the sums. Doing this, one finds schematically
\begin{equation}
F_{\mathbb{S}^3_{\varepsilon}}^{s (k)}(0) \rightarrow  \sum_{n}^{\infty} \sum_{q} m_n^{(s)} \left.\left[ \frac{d^k  }{d \varepsilon^k} \log \lambda^{(s)}_{n,q}(\varepsilon)\right]\right|_{\varepsilon=0} \, ,
\end{equation} 
up to obvious details in each case. We find that for general values of $k$, the sums over $q$ can always be performed once we have evaluated the resulting general terms at $\varepsilon=0$. We are then left with the infinite sums over $n$, which are divergent. In order to deal with those, there exist different strategies. One of them consists in performing the sums up to some finite value $r$, and then expand the result around $r\rightarrow \infty$, extracting the constant piece. 
One may worry that the universal contribution may appear polluted by spurious additional constants which depend on the regularization. In order to isolate those, one possibility is to redefine the summation index $n \rightarrow l\equiv n+k$ for some fixed $k$, replacing the lower limit of the sum accordingly ---\eg if we have $\sum_{n=0}^{\infty}g(n)$, we can consider instead $\sum_{l=j}^{\infty} g(l-j)$. Those constants which fluctuate as we change $j$ cannot be universal, while those which do not, do have chance. In some cases, the resulting sums over $n$ can be written as linear combinations of Riemann's zeta functions, whose analytic continuations can then be used to deal with the divergent parts and produce finite answers. 

2. A second alternative consists in taking derivatives of  $ \myZ'(0)$ ---as defined in terms of the spectral zeta function in \req{spectralzeta}--- with respect to $\varepsilon$ and then evaluating this quantity for $\varepsilon=0$ using the expression (\ref{gpp}) in terms of the heat kernel. In order to do this in practice, we need to evaluate the divergent terms appearing in the heat kernel for small $t$,\footnote{In the cases at hand these can be obtained via the Euler-Maclaurin formula.} plug the resulting expression in  (\ref{gpp}) and then evaluate for $s\rightarrow 0$.  

3. Finally, we can actually use full-fledged numerical results for $F^s_{\mathbb{S}^3_{\varepsilon}}$ and extract the derivatives $F_{\mathbb{S}^3_{\varepsilon}}^{s (k)}(0)$ within the precision allowed by the numerics. This was the method followed in \cite{Bueno:2018yzo} to obtain $F^{(3)}_{\mathbb{S}^3_{\varepsilon}}(0)$ for the scalar and the fermion.

\subsubsection{Conformally-coupled scalar}
Let us consider first the conformally-coupled scalar. Using the first method described above, we find (we omit the ``$s=0$'' label to avoid the clutter in the following expressions)
\begin{align}
F^{(1)}_{\mathbb{S}^3_{\varepsilon}}(0)&= +\frac{1}{2}\sum_{n=0}^{\infty} (n+1) \sum_{q=0}^{n} \frac{\lambda'_{n,q}(0)}{ \lambda_{n,q}(0)}\\ &=-\frac{1}{12}\sum_{n=0}^{\infty} (n+1)^2  \, , \\ 
F^{(2)}_{\mathbb{S}^3_{\varepsilon}}(0)&= +\frac{1}{2}\sum_{n=0}^{\infty} (n+1) \sum_{q=0}^{n} \left[\frac{\lambda''_{n,q}(0)}{ \lambda_{n,q}(0)} - \frac{\lambda'_{n,q}(0)^2}{ \lambda_{n,q}(0)^2} \right]\\ &=+\frac{1}{60}\sum_{n=0}^{\infty}  \frac{(n+1)^2(-15+4n(2+n)(36+28n(2+n)))}{(3+4n(2+n))^2}\, ,\\ 
F^{(3)}_{\mathbb{S}^3_{\varepsilon}}(0)&=+\frac{1}{2}\sum_{n=0}^{\infty} (n+1) \sum_{q=0}^{n} \left[\frac{\lambda'''_{n,q}(0)}{ \lambda_{n,q}(0)} - \frac{3\lambda'_{n,q}(0)\lambda''_{n,q}(0) }{ \lambda_{n,q}(0)^2}+\frac{2\lambda'_{n,q}(0)^3}{\lambda_{n,q}(0)^3} \right] \\
F^{(3)}_{\mathbb{S}^3_{\varepsilon}}(0)&=\frac{-1}{210} \sum_{n=0}^{\infty} \frac{(n+1)^2(105+4n(2+n)(2687+12n(2+n)(165+76n(2+n))))}{(3+4n(2+n))^3}\, ,
\end{align}
and so on. The same sums are obtained using the heat-kernel as defined in \req{fsheat}. As anticipated, all the above expressions are divergent and require some treatment. For instance, the first derivative can be readily rewritten in terms of Riemann's zeta as
\begin{equation}
F^{(1)}_{\mathbb{S}^3_{\varepsilon}}(0)=-\frac{1}{12} \zeta(-2)\, .
\end{equation}
The argument $-2$ corresponds to a ``trivial'' zero of the zeta function, so this evaluates to $F^{(1)}_{\mathbb{S}^3_{\varepsilon}}(0)=0$, as expected. Performing the sum up to some value $r$ for different redefinitions of the summation index $n$ and expanding around $r\rightarrow \infty$ ---as described in the previous subsection--- we can extract the corresponding universal contributions analytically. The results appear summarized in Table \ref{tabli3} up to $k=5$. In particular, we obtain
\begin{equation} \label{f2f3scal}
F^{(2)}_{\mathbb{S}^3_{\varepsilon}}(0)=-\frac{\pi^2}{32}\, , \quad F^{(3)}_{\mathbb{S}^3_{\varepsilon}}(0)=\frac{\pi^2}{1680}\, .
\end{equation}
Now, in our conventions,  \cite{Osborn:1993cr,Buchel:2009sk}
\begin{equation}
\ctt=\frac{3}{32\pi^2} \, , \quad t_4= +4\, ,
\end{equation}
for a conformally-coupled scalar in $d=3$, so one immediately verifies that both \req{f20} and the conjectural relation \req{3conj} are exactly satisfied.

\bgroup
\def\arraystretch{1.8}
\begin{table*}[t]  \hspace{-1.1cm}
	\begin{tabular}{|c|c|c|c|c|c|c|}
		\hline
		& $F_{\mathbb{S}^3_{\varepsilon}}^{(1)}(0)$ & $F_{\mathbb{S}^3_{\varepsilon}}^{(2)}(0)$   & $F_{\mathbb{S}^3_{\varepsilon}}^{(3)}(0)$   &  $F_{\mathbb{S}^3_{\varepsilon}}^{(4)}(0)$  & $F_{\mathbb{S}^3_{\varepsilon}}^{(5)}(0)$ &  $F_{\mathbb{S}^3_{\varepsilon}}^{(6)}(0)$ \\
		\hline\hline
		Scalar &0 & $-\frac{\pi^2}{32}$    &  $\frac{\pi^2}{1680}$ & $\frac{6287\pi^2}{560}-\frac{73\pi^4}{64}$ & $\frac{962567\pi^2}{924}-\frac{390059\pi^4}{3696}$ &  \tiny{$\frac{198003836 \pi ^2}{1001}-\frac{134266337 \pi ^4}{16016}-\frac{37801 \pi^6}{32}$} \\ \hline
		Fermion &0	& $-\frac{\pi^2}{16}$     & $-\frac{\pi^2}{840}$ & $\frac{6313\pi^2}{280}-\frac{73\pi^4}{32}$ & $\frac{963973 \pi^2}{462}-\frac{390721\pi^4}{1848}$ & \tiny{$\frac{396153698 \pi ^2}{1001}-\frac{134382043 \pi ^4}{8008}-\frac{37801 \pi^6}{16}$}  \\ \hline
		\end{tabular}
	\caption{Values of derivatives of $F_{\mathbb{S}^3_{\varepsilon}}$ with respect to the squashing parameter $\varepsilon$ evaluated at $\varepsilon=0$ for a conformally-coupled scalar and a free fermion. }
	\label{tabli3}
\end{table*}
\egroup

The fact that the results obtained using this method are consistent with the general expectations for  $F^{(1)}_{\mathbb{S}^3_{\varepsilon}}(0)$ and $F^{(2)}_{\mathbb{S}^3_{\varepsilon}}(0)$ and produce a value of $F^{(3)}_{\mathbb{S}^3_{\varepsilon}}(0)$ in agreement with our conjecture gives us confidence that the method works well also for higher-order derivatives. A sanity check can be nonetheless performed using the second method described above. 
To proceed we must identify the divergent terms in the small $t$ behavior of the heat kernel. These can be obtained via the Euler-Maclaurin formula and we find that
\be 
K(t) = \frac{\sqrt{\pi} \sqrt{1 + \varepsilon}}{4} \left[\frac{1}{t^{3/2}} - \frac{\varepsilon - 3}{12 \sqrt{t}} + \frac{(15 - 10 \varepsilon + 87 \varepsilon^2) \sqrt{t}}{480} \right] + \mathcal{O}(t)\, .
\ee
The free energy is given by ---see discussion around \req{gsss} above,
\be 
F_{\mathbb{S}^3_\varepsilon} 
= - \frac{1}{2} \zeta_{(0)}'(0) = - \frac{1}{2} G(0) \, , 
\ee
where the subindex in the spectral zeta function refers to the spin of the conformally-coupled scalar field.
To obtain the derivatives of the free energy with respect to the squashing parameter we simply differentiate the spectral zeta function with respect to $\varepsilon$ and evaluate the resulting contributions. For example, for the first derivative after some simple algebraic manipulations ---including summing over the multiplicities of the eigenvalues--- we obtain
\be 
F_{\mathbb{S}^3_\varepsilon}^{(1)}(0) =  \frac{-1}{2} \lim_{t\to 0} \, \left\{  \sum_{n = 0}^\infty \frac{(n + 1)^2}{3} \exp\left[ - \frac{t (2 n + 3)(2 n + 1)}{4} \right] - \frac{\sqrt{\pi}}{12} \left[\frac{1}{t^{3/2}} + \frac{1}{4 \sqrt{t}} + \frac{\sqrt{t}}{32} \right] \right\}\, .
\ee
For any finite value of $t$ the sum converges rapidly provided one goes to large enough (but finite) $n$. Obtaining the value of the sum when $t = 0$ would require including the full infinity of terms. However, there are a couple of ways by which we can extract this relevant part. We will illustrate some of these considerations for $F_{\mathbb{S}^3_\varepsilon}^{(1)}(0)$ but note that they have to be adjusted accordingly for each derivative. 
First let us note that, by analyzing the large-$n$ limit of the sum we see that the terms behave as 
$ 
e^{-(3 + 2 n) t} + \mathcal{O} (n^{-1}) $.
This means that the large $n$ terms have magnitude of  $10^{-a}$ when (assuming $t$ is small) 
$
n_{\rm max} \sim a \ln 10/(2 t) 
$.
Next, note that viewed as a function of $t$ the sum behaves as $\text{sum} = \text{result} + \mathcal{O}(t^\#)$ and thus we can obtain more accurate approximations to the result by including many terms in the sum and analyzing its behavior as a function of $t$ for small $t$.\footnote{In the case of the first derivative, we can also use the Euler-Maclaurin formula to deduce that the first correction behaves like $t^{3/2}$, allowing for a more careful check of the convergence. However, for the higher-order derivatives, the integrals required in the Euler-Maclaurin formula cannot be obtained in closed form, and therefore this double check is less useful in those cases.} As we make $t$ smaller, we will obtain more correct digits in the value of the result. For example, in the present case, we find:
$F_{\mathbb{S}^3_\varepsilon}^{(1)}(0) = -1.923247536 \times 10^{-10}$ with $t = 10^{-4} $, $ n_{\rm max} = 10^5$; $F_{\mathbb{S}^3_\varepsilon}^{(1)}(0) = -1.923235636 \times 10^{-13}$ with $t = 10^{-6} $, $ n_{\rm max} = 10^5$ and $F_{\mathbb{S}^3_\varepsilon}^{(1)}(0) = -1.923235517 \times 10^{-16}$ with $t = 10^{-8} $, $ n_{\rm max} = 10^5$.
The result is consistent with $ F_{\mathbb{S}^3_\varepsilon}^{(1)}(0) = 0$, which we know to be true on general grounds, and we also obtained using the first method.

We present the results for the higher-order derivatives of the free energy with less detailed discussion, but the general idea of the analysis is the same. We have
\begin{align}
F_{\mathbb{S}^3_\varepsilon}^{(2)}(0) =& \frac{1}{2} \lim_{t\to 0} \, \bigg\{\sum_{n = 0}^\infty \bigg[ \frac{(n+1)^2( 112 n^4 + 448 n^3 + 592 n^2 + 288 n - 15)}{15 (2 n + 3)^2 (2n + 1)^2} 
\nonumber\\
&-\frac{t (n + 1)^2 (16n^4+64n^3+56n^2-16n+5)}{20 (2n + 3)(2 n + 1)}  \bigg]\exp \left[ -\frac{ t (2 n + 3)(2 n + 1)}{4} \right] 
\nonumber\\
&- \frac{\sqrt{\pi}}{24} \left[\frac{1}{t^{3/2}} +{\frac {7}{4 \sqrt {t}}}  + {\frac {641\,\sqrt {t}}{160}}\right]
\bigg\} \, ,
\\
F_{\mathbb{S}^3_\varepsilon}^{(3)}(0) =& +\frac{1}{2} \lim_{t\to 0} \, \bigg\{\sum_{n = 0}^\infty \bigg[ \frac{-2(3648n^6+21888n^5+51696n^4+60864n^3+42428n^2+21496n+105)}{(2n + 1)^3(2n + 3)^3}  
\nonumber\\
&+ \frac{t (3072n^6+18432n^5+39024n^4+33216n^3+1672n^2-13936n-105)}{210(2 n + 3)^2(2n + 1)^2}
\nonumber\\
&- \frac{t^2(960n^6+5760n^5+8688n^4-3648n^3-7132n^2+8392n+105)}{1680(2n + 3)(2 n + 1)} 
\bigg]
\nonumber\\
&\times (n+1)^2 \exp \left[ -\frac{ t (2 n + 3)(2 n + 1)}{4} \right] + \frac{\sqrt{\pi}}{16} \left[\frac{1}{t^{3/2}} + \frac{5}{4 \sqrt{t}} - \frac{731 \sqrt{t}}{160} \right]  \bigg \} \, .
\end{align}
Evaluating each of these sums with the same method illustrated for the first derivative, we obtain numerical results completely consistent with the ones obtained using the first method and presented in \req{f2f3scal}.
We have also confirmed the values of the fourth and fifth derivatives presented in Table \ref{tabli3}, although the corresponding sums for those terms are more complicated than they are illuminating and so we do not present them here.



\subsubsection{Free fermion}
Let us now consider the free fermion. Using the first method we find,
 \begin{align}
F^{(1)}_{\mathbb{S}^3_{\varepsilon}}(0)=&\frac{1}{6}\left[ \sum_{n=1}^{\infty} n(n+1) + \sum_{n=2}^{\infty} n(n-1) \right]  \, , \\ 
F^{(2)}_{\mathbb{S}^3_{\varepsilon}}(0)=&\frac{-1}{30} \left[ \sum_{n=1}^{\infty}\frac{(8+32n+25n^2+45n^3+72n^4+28n^5)}{n(1+2n)^2}\right. \\ \notag & \quad \quad \quad \left.+ \sum_{n=2}^{\infty}  \frac{(-8+32n-25n^2+45n^3-72n^4+28n^5)}{n(1-2n)^2} \right]\, ,\\ \notag
F^{(3)}_{\mathbb{S}^3_{\varepsilon}}(0)=&\frac{1}{105} \left[ \sum_{n=1}^{\infty}\frac{(120+720n+1280n^2-581n^4+525n^5+1350n^6+1380n^7+456n^8)}{n^3(1+2n)^3}\right. \\    & \quad  \left.+ \sum_{n=2}^{\infty} \frac{(120-720n+1280n^2-581n^4-525n^5+1350n^6-1380n^7+456n^8)}{n^3(-1+2n)^3} \right]\, . \end{align}
Just like for the scalar, the same sums are obtained using the heat-kernel as defined in \req{fsheat}. Using the first method described above, we obtain $F^{(1)}_{\mathbb{S}^3_{\varepsilon}}(0)=0$. In the case of the fermion, we have  \cite{Osborn:1993cr,Buchel:2009sk}
\begin{equation}
\ctt=\frac{3}{16\pi^2} \, , \quad t_4= -4\, ,
\end{equation}
so based on \req{f20} and \req{3conj} we expect
\begin{equation}
F^{(2)}_{\mathbb{S}^3_{\varepsilon}}(0)=-\frac{\pi^2}{16}\, , \quad F^{(3)}_{\mathbb{S}^3_{\varepsilon}}(0)=-\frac{\pi^2}{840}\, .
\end{equation}
From this, we exactly obtain the expected result for $F^{(2)}_{\mathbb{S}^3_{\varepsilon}}(0)$. On the other hand, for $F^{(3)}_{\mathbb{S}^3_{\varepsilon}}(0)$ we find that the above result appears polluted by an additional constant which does not disappear as we make redefinitions of the summation index. A similar phenomenon occurs for the higher-order derivatives. Hence, carrying out the zeta function regularization method in this case becomes particularly useful.   

In the case of the fermion,
since the eigenvalues come in two branches, we must define spectral zeta functions for each branch, which we denote as $\zeta_+$ and $\zeta_-$. The zeta function for the squared Dirac operator, from which the effective action is derived, is then
\be 
\zeta_{1/2}(s) = \zeta_+(2s) + \zeta_-(2s) \, .
\ee
The analytic continuation of $\zeta_{1/2}(s) $ then proceeds in exactly the same way as above and we have
\be 
F_{\mathbb{S}^3_\varepsilon} = - \frac{1}{2} \zeta_{1/2}'(0) = - \frac{1}{2} G(0) \, .
\ee
The basic method of evaluation is mostly identical. We require the divergent expansion of the heat kernel which in this case takes the form
\be 
K(t) = \frac{\sqrt{\pi(1+\varepsilon)}}{32} \left[ \frac{1}{t^{3/2}} +  \frac{2(\varepsilon - 3)}{12 \sqrt{t}} \right] + \mathcal{O}(t^{3/2}) \, .
\ee
We then obtain (convergent) sums in the same manner as for the scalar. However, in this case we find that the convergence is less rapid than for the scalar. To overcome this difficulty, we evaluate the relevant sums for several values of $t$ ranging from $t = 10^{-10}$ to $t = 10^{-4}$, fit the resulting data to a form 
\be 
\frac{d^k F^{(k)}_{\mathbb{S}^3_\varepsilon}(t, \varepsilon)}{d \varepsilon^k}  \bigg|_{\varepsilon = 0} =  F^{(k)}_{\mathbb{S}^3_\varepsilon}(0) + \sum_i b_{i} t^{i/2} \, ,
\ee
and then extract the $t \to 0$ behavior from the fit. This allows us to  reach precisions of order $10^{-10}$ or better. Here, for these evaluations, we include up to $5 \times 10^{6}$ terms in the sums, which guarantees strong convergence for these values of $t$.

Now, let us consider explicitly the first three derivatives. For simplicity of presentation we have combined the sums for the positive and negative branches by redefining the summation index for the latter by $n \to n+1$ (this is perfectly justified  since the sum is convergent for any finite $t$). We have:
\begin{align}
F_{\mathbb{S}^3_\varepsilon}^{(1)}(0) =& -\frac{1}{2} \lim_{t\to 0} \, \bigg\{ - \frac{2}{3} \sum_{n=1}^\infty  n(n+1) \exp \left[-t(1+2n)^2\right] 
 - \frac{\sqrt{\pi}}{48} \left[- \frac{1}{t^{3/2}} + \frac{2}{\sqrt{t}} \right]  \bigg\}\, ,
\nonumber\\
F_{\mathbb{S}^3_\varepsilon}^{(2)}(0) =& -\frac{1}{2} \lim_{t\to 0} \, \bigg\{ \frac{2}{15} \sum_{n=1}^\infty \bigg( \frac{(28 n^6+84 n^5+77 n^4+14 n^3+13 n^2+20 n+4)}{n(n+1)(1+2n)^2} 
\nonumber\\
&- t (12 n^4 + 24 n^3 + 11 n^2 - n - 16) \bigg) \exp \left[-t(1+2n)^2\right] - \frac{\sqrt{\pi}}{96} \left[\frac{1}{t^{3/2}} - \frac{14}{\sqrt{t}} \right] \bigg\} \, ,
\nonumber\\
F_{\mathbb{S}^3_\varepsilon}^{(3)}(0) =& -\frac{1}{2} \lim_{t\to 0} \, \bigg\{ \frac{2}{105} \sum_{n=1}^\infty \bigg(\frac{1}{n^3(n+1)^3(1+2n)^2} \big( -456 n^{10}-2280 n^9-4518 n^8-4392 n^7
\nonumber\\
&-1348 n^6+1752 n^5+802 n^4-1880 n^3-2120 n^2-840 n-120 \big) 
\nonumber\\
&+ \frac{2 t (192 n^8+768 n^7+1017 n^6+363 n^5+145 n^4+581 n^3-324 n^2-682 n-240)}{n^2 (n+1)^2}
\nonumber\\
&- \frac{t^2 (1+2n)^2 (60 n^6+180 n^5+87 n^4-126 n^3-397 n^2-304 n+640 )}{n(n+1)}  \bigg)\nonumber\\
& \times  \exp \left[-t(1+2n)^2\right]
- \frac{\sqrt{\pi}}{64} \left[-\frac{1}{t^{3/2}} + \frac{10}{\sqrt{t}} \right] \bigg \}\, .
\end{align}
Using the analysis described above, we find the following fits: 
\begin{align} 
 F_{\mathbb{S}^3_\varepsilon}^{(1)}(0)  &= -{ 1.547434165\times 10^{-12}}+ 0.0000001115593597\,\sqrt {t}-
 0.001285814355\,t+ 3.0\,{t}^{3/2}\, ,
\nonumber\\
 F_{\mathbb{S}^3_\varepsilon}^{(2)}(0) &= - 0.6168502751+ 0.9453088555\,\sqrt {t}- 0.001555923570\,t+ 3.0\,{t}^{
3/2}
 \, ,
\nonumber\\
F_{\mathbb{S}^3_\varepsilon}^{(3)}(0) &=- 0.01174952914+ 1.417966266\,\sqrt {t}- 0.01252516826\,t+ 3.0\,{t}^{3
/2}\, ,
\end{align}
where here we have included coefficients with 10 digits of precision but in the actual computations we have worked to 100 digits. The results for  $ F_{\mathbb{S}^3_\varepsilon}^{(1)}(0) $ and $ F_{\mathbb{S}^3_\varepsilon}^{(2)}(0)$ are consistent with the expected values, whereas the one for  $ F_{\mathbb{S}^3_\varepsilon}^{(3)}(0) $ exactly agrees with the expectation based on our conjectural relation in \req{3conj} (which using the first method we had obtained analytically but polluted with an additional spurious constant). 
Proceeding similarly with the higher-order derivatives we can identify which of the constants obtained analytically using the first method are universal and which ones are spurious. The final values appear presented in Table \ref{tabli3}.



\subsection{Five-dimensional CFTs}
Let us now move to five dimensions. In this case we will restrict ourselves to the case of the conformally-coupled scalar.

\subsubsection{Conformally-coupled scalar}
 The eigenvalues and multiplicities of the conformal Laplace operator on $\mathbb{S}^5_{\varepsilon}$ read \cite{Bobev:2017asb}
%
%
\begin{align}
\lambda^{(0)}_{n,q}(\varepsilon)&=(n-1)(n+3)-\frac{\varepsilon(n-1-2q)^2}{(1+\varepsilon)}+\frac{3}{4}(5-\varepsilon)\, ,\label{eigensq5d}\\
m_{n,q}^{(0)}&=\frac{\left(n+1\right)(q+1)(n-q)}{2}\, ,\label{degesq5d}
\end{align}
where the integers $n$ and $q$ obey $n\geq1$ and $0\leq q\leq n-1$ respectively. Therefore, we can write the free energy as
\begin{equation}
F^{0}_{\mathbb{S}^5_{\varepsilon}}= +\frac{1}{2}\sum_{n=1}^{\infty} \sum_{q=0}^{n-1} m_{n,q}^{(0)} \log \lambda^{(0)}_{n,q}(\varepsilon) \, .
\end{equation}
Proceeding analogously to three-dimensional scalar and fermion cases, we find using the first method described above
\begin{align}
F^{(1)}_{\mathbb{S}^3_{\varepsilon}}(0)=& \sum_{n=1}^{\infty} \frac{-n(1+n)^2(2+n)}{120}   \, , \\ 
F^{(2)}_{\mathbb{S}^3_{\varepsilon}}(0)=&  \sum_{n=1}^{\infty} \frac{n(1+n)^2(2+n)(-1131+16n(2+n)(-16+11n(2+n)))}{840 (3+4n(2+n))^2} \, , \\
F^{(3)}_{\mathbb{S}^3_{\varepsilon}}(0)=&  \sum_{n=1}^{\infty} \frac{-n(1+n)^2(2+n)(-54081+4n(2+n)(-1509+4n(2+n)  (29+164n(2+n)) ))}{1260 (3+4n(2+n))^3} \, , 
 \end{align}
and similar expressions for the following higher derivatives. Summing the above expressions up to some finite value $r$ and expanding around $r\rightarrow \infty$ for various redefinitions of the summation index $n$, we obtain the following universal contributions
\begin{equation} \label{f5k}
F^{(1)}_{\mathbb{S}^5_{\varepsilon}}=0\, , \quad\quad F^{(2)}_{\mathbb{S}^5_{\varepsilon}}=\frac{3\pi^2}{256}\, , \quad  \quad F^{(3)}_{\mathbb{S}^5_{\varepsilon}}=\frac{151\pi^2}{4480}\, ,
\end{equation}  
which appear in Table \ref{tabli5} along with the $k=4,5,6$ derivatives. The values of $F^{(1)}_{\mathbb{S}^5_{\varepsilon}}$ and $F^{(2)}_{\mathbb{S}^5_{\varepsilon}}$ found agree with the general expectations, whereas the one obtained for $F^{(3)}_{\mathbb{S}^5_{\varepsilon}}$ precisely agrees with the prediction following from our new conjectural relation \req{5conj2}, as can be readily verified using the known values \cite{Osborn:1993cr,Buchel:2009sk}
\begin{equation}
\ctt=\frac{45}{256\pi^4} \, , \quad t_2=0 \, , \quad t_4=+12\, ,
\end{equation}
corresponding to a conformal scalar in $d=5$. In this case, all spurious constants can be removed as described above. Nonetheless, just like for the $d=3$ scalar, it is good to perform alternative checks of the values presented in Table \ref{tabli5}. 

\bgroup
\def\arraystretch{1.8}
\begin{table*}[t] \hspace{-1.65cm}
	\begin{tabular}{|c|c|c|c|c|c|c|}
		\hline
		& $F_{\mathbb{S}^5_{\varepsilon}}^{(1)}(0)$ & $F_{\mathbb{S}^5_{\varepsilon}}^{(2)}(0)$   & $F_{\mathbb{S}^5_{\varepsilon}}^{(3)}(0)$   &  $F_{\mathbb{S}^5_{\varepsilon}}^{(4)}(0)$  & $F_{\mathbb{S}^5_{\varepsilon}}^{(5)}(0)$  & $F_{\mathbb{S}^5_{\varepsilon}}^{(6)}(0)$ \\
		\hline\hline
		Scalar &0 & $\frac{3\pi^2}{256}$    &  $\frac{151\pi^2}{4480}$ & $ \frac{147 \pi^4}{128} - \frac{2235553 \pi^2}{197120}$  &\tiny{ $\frac{104026563 \pi ^4}{640640}-\frac{51334449 \pi ^2}{32032}$} & \tiny{$\frac{-639254581807 \pi ^2}{1537536}+\frac{27998928053 \pi ^4}{1537536}+\frac{77541\pi ^6}{32}$} \\ \hline
		\end{tabular}
	\caption{Values of derivatives of $F_{\mathbb{S}^5_{\varepsilon}}$ with respect to the squashing parameter $\varepsilon$ evaluated at $\varepsilon=0$ for a conformally-coupled scalar. }
	\label{tabli5}
\end{table*}
\egroup

In order to do that, let us first consider the second method described above. 
Again, we must identify the divergent terms in the small $t$ behavior of the heat kernel. These can be obtained via the Euler-Maclaurin formula and we find that
\begin{align}\notag
K(t) =& \frac{\sqrt{\pi} \sqrt{1 + \varepsilon}}{32} \left[\frac{1}{t^{5/2}} + \frac{\varepsilon-5}{12 t^{3/2}} + \frac{279 \varepsilon^2 + 26 \varepsilon - 65}{480 \sqrt{t}} - \frac{\sqrt{t} (31733 \varepsilon^3 + 16293 \varepsilon^2 - 441 \varepsilon + 735)}{40320} \right]        \\� &+ \mathcal{O}(t) \, .
\end{align}
We can then proceed exactly as in the case of the $d = 3$ scalar. 
The first derivative reduces to
\begin{align} 
F_{\mathbb{S}^5_\varepsilon}^{(1)}(0) =&  \frac{1}{2} \lim_{t\to 0} \, \bigg\{  \sum_{n = 0}^\infty -\frac{(n+1)(n + 2)^2(n+3)}{60} \exp\left[ - \frac{t (2 n + 5)(2 n + 3)}{4} \right] 
\nonumber\\
&- \frac{\sqrt{\pi}}{160} \left[-\frac{1}{t^{5/2}} + \frac{5}{12 t^{3/2}} +  \frac{13}{96 \sqrt{t}} + \frac{7\sqrt{t}}{384} \right] \bigg\}\, .
\end{align}
Evaluating this sum in the way described for the $d = 3$ scalar, we find: $F_{\mathbb{S}^5_\varepsilon}^{(1)}(0) = 8.714716523\times10^{-12}$ with $t = 10^{-4} $ , $n_{\rm max} = 10^5$; $F_{\mathbb{S}^5_\varepsilon}^{(1)}(0) = 8.714661485\times10^{-15}$ with $t = 10^{-6} $ , $n_{\rm max} = 10^5$ and  $F_{\mathbb{S}^5_\varepsilon}^{(1)}(0) =8.714676591 \times10^{-18}$ with $t = 10^{-8} $ , $n_{\rm max} = 10^5$. 
The result is consistent with $F_{\mathbb{S}^5_\varepsilon}^{(1)}(0) = 0$, as it must be, and is completely independent of any spurious constant. 

For the second derivative we obtain 
\begin{align} 
F_{\mathbb{S}^5_\varepsilon}^{(2)}(0) =& \frac{1}{2} \lim_{t\to 0} \, \bigg\{  \sum_{n = 0}^\infty \bigg(\frac{176 n^4 + 1408 n^3 + 3616 n^2 + 3200 n - 315}{420(2n+5)(2n+3)} 
\nonumber\\
&- \frac{t}{1680} (48n^4 + 384 n^3 + 808 n^2 + 160 n + 315) \bigg) \frac{(n+1) (n+2)^2 (n+3)}{(2n + 5)(2n+3)}  
\nonumber\\
&\times \exp\left[ - \frac{t (2 n + 5)(2 n + 3)}{4} \right] - \frac{\sqrt{\pi}}{320} \left[\frac{1}{t^{5/2}} - \frac{5}{4 t^{3/2}} - \frac{2401}{96 \sqrt{t}} - \frac{14205 \sqrt{t}}{896} \right] \bigg\}
\end{align}
Evaluating this expression we find
\begin{align}
F_{\mathbb{S}^5_\varepsilon}^{(2)}(0) =& 0.1168817449 \quad \text{with} \quad t = 10^{-4} \, , \quad n_{\rm max} = 10^5 \, ,
\\
F_{\mathbb{S}^5_\varepsilon}^{(2)}(0) =&   0.1156594266 \quad \text{with} \quad t = 10^{-6} \, , \quad n_{\rm max} = 10^5 \, ,
\\
F_{\mathbb{S}^5_\varepsilon}^{(2)}(0) =&   0.1156594266 \quad \text{with} \quad t = 10^{-8} \, , \quad n_{\rm max} = 10^5 \, .
\end{align}
We see that already at $t= 10^{-6}$ the numerical evaluation of the sum agrees with the ``predicted value'' appearing in  \req{f5k}
to more than 10 decimal places. At $t = 10^{-8}$ the agreement holds up to 15 decimal places. 

Finally let us present the relevant sum for determining the third derivative:
\begin{align}
F_{\mathbb{S}^5_\varepsilon}^{(3)}(0) =& \frac{1}{2} \lim_{t\to 0} \, \bigg\{  \sum_{n = 0}^\infty \nonumber\\ & \hspace{-0.5cm} \bigg[-\frac{2624n^6+31488n^5+150032n^4+360576n^3+452876n^2+270384n+2835}{630 (2n+5)^2(2n+3)^2} 
\nonumber\\
&+ \frac{t(1408n^6+16896n^5+73744n^4+139392n^3+87664n^2-43584n-2835)}{2520 (2 n +5)(2n + 3)}  
\nonumber\\
&- \frac{t^2}{20160}  (320n^6+3840n^5+14096n^4+10368n^3-15268n^2+19824n+2835 )\bigg]
\nonumber\\
&\times \frac{(n+1) (n+2)^2 (n+3)}{(2n + 5)(2n+3)}\exp\left[ - \frac{t (2 n + 5)(2 n + 3)}{4} \right]
\nonumber\\
&- \frac{3\sqrt{\pi}}{256} \left[\frac{1}{t^{5/2}} - \frac{7}{12 t^{3/2}} + \frac{141}{32 \sqrt{t}} - \frac{639689 \sqrt{t}}{40320} \right] \bigg\}
\end{align}
This yields
\begin{align}
F_{\mathbb{S}^5_\varepsilon}^{(3)}(0) =& 0.3326586184 \quad \text{with} \quad t = 10^{-4} \, , \quad n_{\rm max} = 10^5 \, ,
\\
F_{\mathbb{S}^5_\varepsilon}^{(3)}(0) =&   0.3326585413 \quad \text{with} \quad t = 10^{-6} \, , \quad n_{\rm max} = 10^5 \, ,
\\
F_{\mathbb{S}^5_\varepsilon}^{(3)}(0) =&   0.3326585412 \quad \text{with} \quad t = 10^{-8} \, , \quad n_{\rm max} = 10^5 \, .
\end{align}
Again, by $t = 10^{-6}$ the result agrees with the one obtained using the first method appearing in \req{f5k} up
to 10 decimal places. At $t = 10^{-8}$ the agreement holds to 14 decimal places. The result is then consistent with this exact form, and again is free from any spurious constants. 
Higher derivatives can be computed analogously and the results for $k=4,5,6$ agree with the ones presented in Table \ref{tabli5}. 

\begin{figure}
\centering
\includegraphics[width=7.6cm]{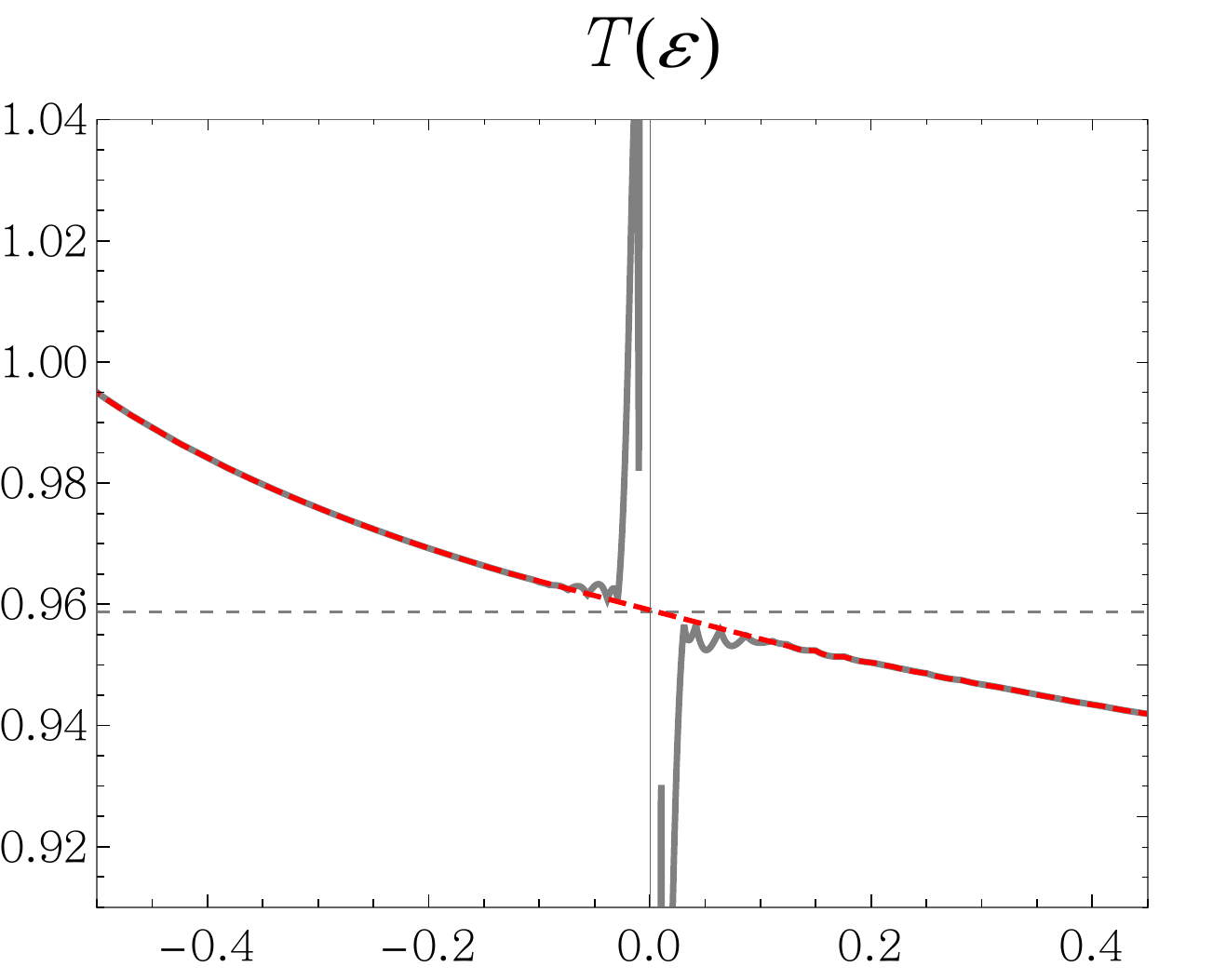}
\includegraphics[width=7.6cm]{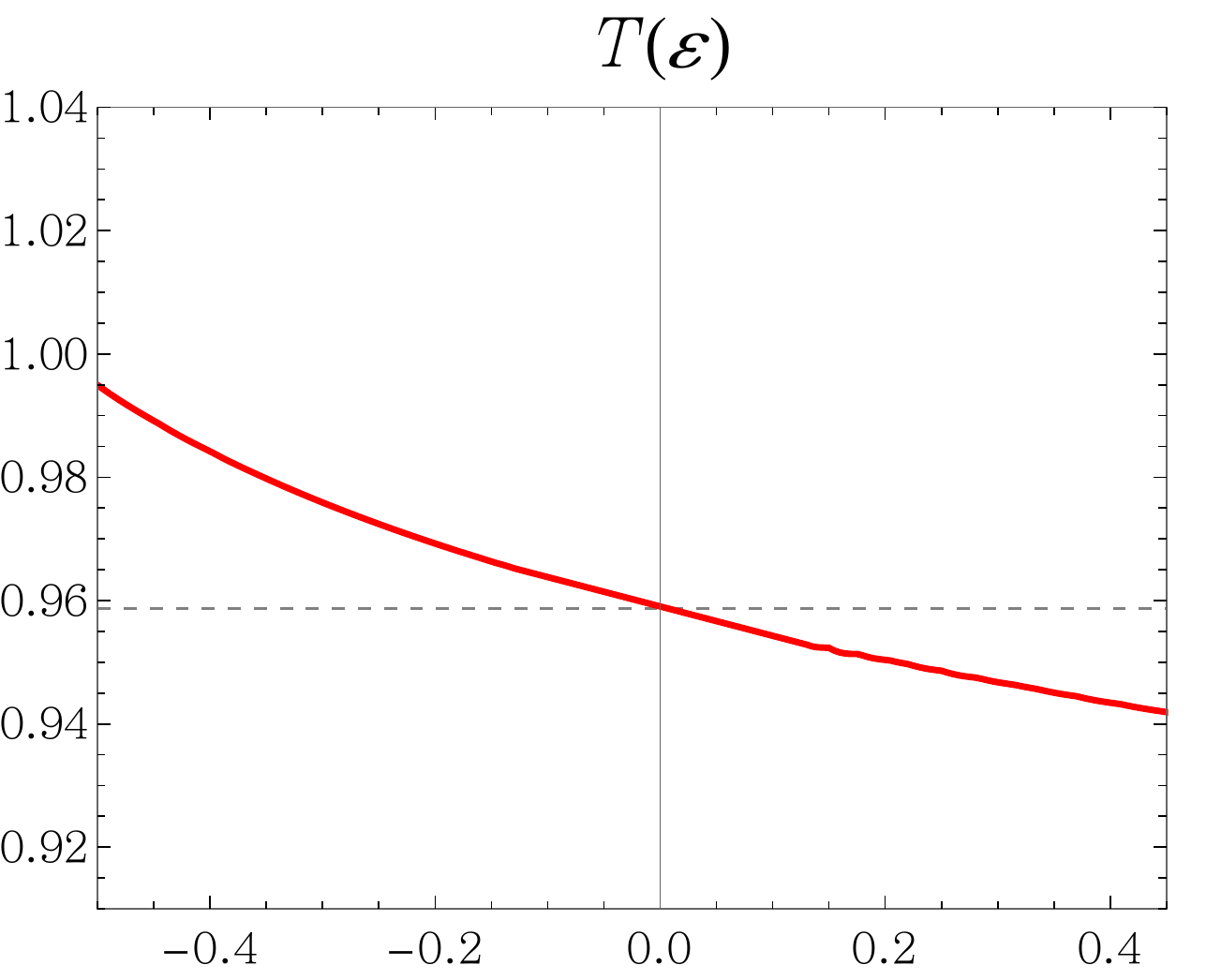}
\caption{$T(\varepsilon)$ using numerical data for  $F_{\mathbb{S}^5_\varepsilon}$  as defined in \req{tep} for values of $\varepsilon$ near $\varepsilon=0$. The horizontal gray dashed line corresponds to the predicted value of $T(0)$ assuming the validity of our conjecture \req{5conj2}. The numerical data is not well behaved very close to $\varepsilon=0$ due to the $1/\varepsilon^3$  and $1/\varepsilon$ powers involved in its definition (see left plot), but the tendency for not so small values of $\varepsilon$ is very neat, and a proper treatment of the data removing the problematic points shows a perfect agreement with the prediction (right). }
\label{5Plot}
\end{figure}

Before closing, let us perform yet another check of the validity of the analytic result found for $F_{\mathbb{S}^5_\varepsilon}^{(3)}(0) $. For that we use the numerical data obtained in \cite{Bobev:2017asb} for $F_{\mathbb{S}^5_\varepsilon}$ and the same method used in \cite{Bobev:2017asb} for the three-dimensional scalar and fermion. The idea is the following. If the conjectural relation in \req{5conj2} holds, the numerical plot of the function
\begin{equation}\label{tep}
T(\varepsilon)\equiv \frac{30 (F_{\mathbb{S}^5_\varepsilon}-F_{\mathbb{S}^5_0}  )}{\ctt \pi^6 \varepsilon^3}- \frac{1}{\varepsilon} \, ,
\end{equation}
should be such that 
\begin{equation}
T(0)=\frac{30}{45} \left[1+\frac{3}{40}t_2+\frac{23}{630}t_4 \right] \, .
\end{equation}
Namely, the function $T(\varepsilon)$ should cross the $\varepsilon=0$ axis at that value.
In the case of the scalar, this yields $T(0)=302/315\simeq 0.95873$. Plotting $T(\varepsilon)$ and the predicted value obtained using our conjecture we again find perfect agreement ---see Fig. \ref{5Plot}.



\section{Conclusion} \labell{discuss}
In this paper we have presented compelling evidence in favor of two conjectures ---summarized in \req{conjs}--- relating the subleading term in the small-squashing expansion of the free energy of squashed-spheres with the stress-tensor three-point function charges $t_2$ and $t_4$ for three- and five-dimensional CFTs respectively.
The evidence in favor of the three-dimensional version ---conjectured originally in \cite{Bueno:2018yzo}--- includes now free scalars and fermions, as well as an infinite family of holographic higher-curvature theories of the GQT class. As for the five-dimensional one, which we have presented here, we have proven it to hold for general QT gravities as well as for quartic and quintic GQT theories admitting Taub-NUT solutions of the form \req{FFnut} and for a conformally-coupled scalar. We  did not consider the case of five-dimensional free fermions, which would provide an additional test of the conjecture. Of course, the next natural step would be to prove both expressions in general using field-theoretical methods, although this looks like a rather challenging computation. 
Finally, it would be interesting to determine whether formulas similar to the ones in \req{conjs} hold for general dimensions, or if, on the contrary, these are related to specific properties of three- and five-dimensional CFTs. 


\section*{Acknowledgments}
We thank Nikolay Bobev, Horacio Casini, Jos\'e Edelstein, Gonzalo Torroba and Yannick Vreys for useful discussions.
 The work of PB was supported by the Simons foundation through the It From Qubit Simons collaboration. PAC was supported by the KU Leuven grant ``Bijzonder Onderzoeksfonds C16/16/005 --- Horizons in hoge-energie fysica''. The work of RAH is supported by the Natural Sciences and Engineering Research Council of Canada through the Banting Postdoctoral Fellowship program. The work of VAP was supported by CONICET. 
 The work of AR was supported by a FPI-UAM pre-doctoral grant.

\appendix

\section{A general-order relation between $h_q$ and $F_{\mathbb{S}_{\varepsilon}^3}$?}\label{hFs}
In this appendix we explore the possibility that $F_{\mathbb{S}^3_{\varepsilon}}$ and $h_q$ are actually related at all orders in the expansions of $\varepsilon$ and $q$, respectively. This is motivated by the GQT gravities results obtained in Section \ref{gqt3} ---see comments at the end of that section. 

Using \req{f1ups} and \req{hqqq}, we can 
%
compare the expansion of $F_{\mathbb{S}^3_{\varepsilon}}$ around $\varepsilon=0$ with the expansion of $h_q$ around $q=1$
\begin{align}
\notag
F_{\mathbb{S}^3_{\varepsilon}}=&F_{\mathbb{S}^3}+\frac{\pi L^2}{G}\Bigg[\frac{\varepsilon ^2 \Upsilon '\left(\chi _0\right)}{2 \chi _0}-\frac{\varepsilon ^3}{6}  \Upsilon ''\left(\chi _0\right)+\frac{\varepsilon ^4 }{24} \left[3 \Upsilon ''\left(\chi _0\right)+\chi _0 \Upsilon
   ^{(3)}\left(\chi _0\right)\right]\\
   &-\frac{\varepsilon ^5}{120}  \left[12 \Upsilon ''\left(\chi _0\right)+\chi _0 \left(8
   \Upsilon ^{(3)}\left(\chi _0\right)+\chi _0 \Upsilon ^{(4)}\left(\chi
   _0\right)\right)\right]+\ldots\Bigg]\, ,\\
   \notag
h_q=&\frac{L^2}{G}\Bigg[-\frac{(q-1) \Upsilon '\left(\chi _0\right)}{8 \chi _0}+\frac{7}{128} (q-1)^2
   \left(\frac{2 \Upsilon '\left(\chi _0\right)}{\chi _0}+\Upsilon ''\left(\chi_0\right)\right)\\\notag
   &\quad \quad +\frac{(q-1)^3}{3072} \left[-\frac{444 \Upsilon '\left(\chi _0\right)}{\chi_0}-156 \Upsilon ''\left(\chi _0\right)+\chi _0 \left(-\frac{27 \Upsilon ''\left(\chi_0\right){}^2}{\Upsilon '\left(\chi _0\right)}-40 \Upsilon ^{(3)}\left(\chi_0\right)\right)\right]\\\notag
   &\quad \quad \, +\frac{(q-1)^4}{98304} \left[\frac{17832 \Upsilon '\left(\chi_0\right)}{\chi _0}+4908 \Upsilon ''\left(\chi _0\right)-\frac{135 \chi _0^2 \Upsilon''\left(\chi _0\right){}^3}{\Upsilon '\left(\chi _0\right){}^2}\right.\\ \notag
   &\quad \quad \quad \, \quad \quad \quad \,\quad  \, \left.+\frac{18 \chi _0\Upsilon ''\left(\chi _0\right) \left(87 \Upsilon ''\left(\chi _0\right)+40 \chi _0\Upsilon ^{(3)}\left(\chi _0\right)\right)}{\Upsilon '\left(\chi _0\right)} \right. \\  & \quad \quad \quad \, \quad \quad \quad \,\quad  \, \left.  +16 \chi _0\left(94 \Upsilon ^{(3)}\left(\chi _0\right)+13 \chi _0 \Upsilon ^{(4)}\left(\chi_0\right)\right)\right]+\ldots\Bigg]\, .
\end{align}
All the terms in both expansions are determined by derivatives of $\Upsilon$ at $\chi=\chi_0$, and the key observation is that all of these derivatives are independent, since $\Upsilon$ is essentially an arbitrary analytic function, with an expansion of the form  $\Upsilon(\chi)=1-\chi+\sum_{n=3}^{\infty}\lambda_n\chi^n$
for any sequence of parameters $\lambda_n$. Therefore, there is a unique correspondence between the derivatives of $\Upsilon$ and the derivatives of $F_{\mathbb{S}^3_{\varepsilon}}$ and, analogously, one involving the derivatives of $\Upsilon$ and those of $h_q$. In turn, this implies a correspondence between derivatives of $F_{\mathbb{S}^3_{\varepsilon}}$ and $h_q$. The first equalities read
\begin{align}\label{Pablonian}
F^{(2)}_{\mathbb{S}^3_{\varepsilon}}(0)=&-8 \pi  h_q'(1)\, ,\\
F^{(3)}_{\mathbb{S}^3_{\varepsilon}}(0)=&-\frac{16\pi}{7} \left[7 h_q'(1)+4 h_q''(1)\right]\, ,\\ \label{Pablonian4}
F^{(4)}_{\mathbb{S}^3_{\varepsilon}}(0)=&+\frac{32\pi }{245}  \left[735 h_q'(1)+126 h_q''(1)+\frac{54 h_q''(1){}^2}{h_q'(1)}-98
   h_q{}^{(3)}(1)\right]\, ,\\ \notag
F^{(5)}_{\mathbb{S}^3_{\varepsilon}}(0)=&-\frac{128 \pi}{4459}  \bigg[40131 h_q'(1)+12054 h_q''(1)+\frac{4536 h_q''(1){}^2}{h_q'(1)}+\frac{1242
   h_q''(1){}^3}{h_q'(1){}^2}\\
   &\quad \quad \quad \quad -3430 h_q{}^{(3)}(1)-\frac{1764 h_q''(1)
   h_q{}^{(3)}(1)}{h_q'(1)}+686 h_q{}^{(4)}(1)\bigg]\, ,
\end{align}
and so on.
These are identities that relate different quantities of a CFT and they hold for an infinite number of holographic higher-order gravities. Therefore, we may suspect that these relations are universal for any CFT. However, as we will now discuss, computations for free fermions cast doubt on the generality of this result.

As shown in~\cite{Hung:2014npa}, the scaling dimension of
 twist operators can be computed from the energy density of the CFT state on the background $\mathbb{S}^1 \times \mathbb{H}^{d-1}$
\be\label{hqEEEEE} 
h_q = \frac{2 \pi q}{d-1} R^{d} \left[\mathcal{E}(T_0) - \mathcal{E}(T_0/q) \right] \, .
\ee
The energy density can be in turn computed from the corresponding partition function. Here we will be concerned only with the $d=3$ Dirac fermion\footnote{Certain subtleties concerning the evaluation of higher-order derivatives of $h_q$ for the conformally coupled scalar were identified in~\cite{Hung:2014npa} and later addressed in~\cite{Lee:2014zaa}.} for which the relevant partition function reads~\cite{Hung:2014npa}
\be 
\log Z^{1/2}(\beta) = \frac{V_\Sigma}{2 \pi^2 \beta R} \int_0^\infty dx \coth \frac{x}{2} \frac{2 \beta^2 \sinh \frac{\pi R x}{\beta} - \pi R x \left(\beta + \pi R x \coth \frac{\pi R x}{\beta} \right)}{x^3 \sinh \frac{\pi R x}{\beta}} \, ,
\ee
where $R^{2} V_\Sigma$ is the (regulated) volume of the $\mathbb{H}^2$. From the partition function the energy density is obtained in the usual way,
\be 
\mathcal{E}(\beta)  = -\frac{1}{R^{2} V_\Sigma} \frac{\partial}{\partial \beta} \log Z^{1/2}(\beta) \, ,
\ee
and then, using \eqref{hqEEEEE}, the derivatives of derivatives of $h_q$ at $q = 1$ can be easily obtained.\footnote{Equivalently, one could have used the expressions appearing \eg in \cite{Bueno:2015qya,Dowker:2015qta}.} By differentiating with respect to $\beta$, substituting $\beta = 2 \pi R$, and then evaluating the (convergent) integral we find 
\be
\frac{\partial^2 \log Z}{\partial \beta^2} \bigg|_{\beta = 2\pi} = \frac{1}{256 \pi} \, , \quad \frac{\partial^3 \log Z}{\partial \beta^3} \bigg|_{\beta = 2\pi}  = \frac{-7}{960 \pi^2} \, , \quad \frac{\partial^4 \log Z}{\partial \beta^4} \bigg|_{\beta = 2\pi}  = \frac{-1}{1024 \pi} + \frac{7}{246 \pi^3} \, ,
\ee
from which we obtain
\be 
h'_q(1) = \frac{\pi}{128} \, , \quad h''_q(1) = -\frac{-13 \pi}{960} \, , \quad h'''_q(1) = \frac{21 \pi }{160} - \frac{\pi^3}{128} \, .
\ee
Using these results in the first two expressions of \eqref{Pablonian} we find agreement with the results obtained for $F^{(2)}_{\mathbb{S}^3_{\varepsilon}}(0)$ and $F^{(3)}_{\mathbb{S}^3_{\varepsilon}}(0)$. However, the prediction for  $F^{(4)}_{\mathbb{S}^3_{\varepsilon}}(0)$ yields
\be 
\left(F^{(4)}_{\mathbb{S}^3_{\varepsilon}}(0)\right)_{\rm prediction} = \frac{\pi^4}{10} - \frac{24189 \pi^2}{24500} \approx -0.003411746 \, .
\ee
This answer disagrees with the result obtained via explicit computation of $F^{(4)}_{\mathbb{S}^3_{\varepsilon}}(0)$ for the free fermion which yields 
\be 
F^{(4)}_{\mathbb{S}^3_{\varepsilon}}(0) = \frac{6313\pi^2}{280}-\frac{73\pi^4}{32}  \approx  0.3098417 \, .
\ee
\begin{figure}
\centering
\includegraphics[scale=0.62]{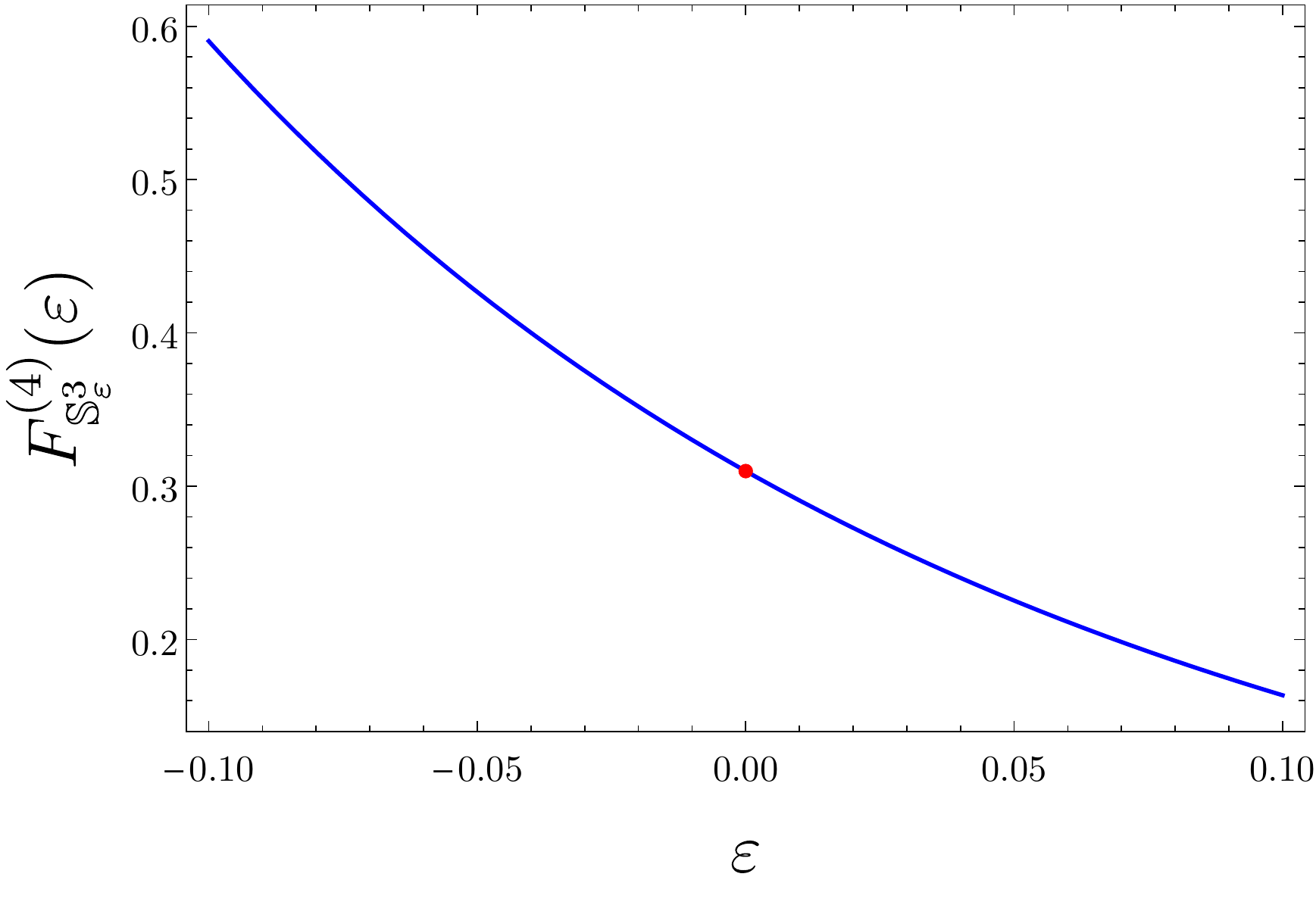}
\caption{Comparison of analytical prediction for $F_{\mathbb{S}^3_\varepsilon}^{(4)}(0)$ of free fermion (red dot) with numerical derivative computed from interpolation of raw numerical data for $F_{\mathbb{S}^3_\varepsilon}(\varepsilon)$.}
\label{fermPlot}
\end{figure}
We have considerable confidence in this exact value for $F^{(4)}_{\mathbb{S}^3_{\varepsilon}}(0)$ --- it can be obtained using the zeta function regularization discussed in the main text, and also from a numerical evaluation of the derivative based on raw numerical data for $F_{\mathbb{S}^3_{\varepsilon}}(\varepsilon)$ --- see Fig.~\ref{fermPlot}. Here the numerical computation of $F_{\mathbb{S}^3_\varepsilon}(\varepsilon)$ was carried out using the zeta function regularization described in the main text. For each value of $\varepsilon$, a total of 3000 terms were included in the sum with the result evaluated for several values of $t$ ranging from $10^{-4}$ to $10^{-7}$. The results of these computations were then fit and the $t \to 0$ behaviour extracted from the fit, giving about 10 digits of precision in the final answer. This procedure was completed for $\varepsilon \in (-1/5, 1/5)$ with a spacing of $10^{-4}$. We then used a 10th-order interpolation of the final results for $F_{\mathbb{S}^3_\varepsilon}(\varepsilon)$ and from this interpolation extracted the derivatives. Due to the precision loss/rounding error in the derivative computation,   the value of the fourth derivative obtained in this way is only trustworthy to about two decimal places. Nonetheless, we see perfect agreement with the analytic result.

Thus, it seems that the general order relationship between $h_q$ and the expansion of $F_{\mathbb{S}^3_\varepsilon}(\varepsilon)$ for small $\varepsilon$ does not hold in general. 

\section{$\Upsilon''(\chi_0)$ and energy-flux parameters}\label{upsit2t4}
%
In the main text we introduced \req{fee0e}, which exactly computes the free energy of a CFT on a squashed sphere in terms of the gravitational Lagrangian for certain types of theories ---namely, those corresponding
 to the special GQT type allowing for single-function Taub-NUT solutions. However, near $\varepsilon=0$, the previous formula has a more universal character, and it predicts the correct value of $F_{\mathbb{S}^d}$, $F_{\mathbb{S}^d_{\varepsilon}}^{(1)}(0)$ and $F_{\mathbb{S}^d_{\varepsilon}}^{(2)}(0)$ for any Einstein-like theory. In particular, $F_{\mathbb{S}^d_{\varepsilon}}^{(2)}(0)\propto \Upsilon'(\chi_0)\propto \ctt$, where the last proportionality applies for any theory of that type. In view of this, its is natural to wonder whether the expression for the third derivative of $F_{\mathbb{S}_{\varepsilon}^{d}}$ in terms of the derivatives of $\Upsilon$ is also universal in some sense. Using \eqref{fee0e}, we find for the third derivative,
 \begin{align} \label{sw3ap} 
F_{\mathbb{S}^d_{\varepsilon}}^{(3)}(0)= \, \frac{(-1)^{\frac{(d+1)}{2}}\pi^{\frac{d}{2}}(d^2-1)L^{d-1}}{16\Gamma[\frac{d}{2}]\chi_0^{\frac{d-1}{2}} G}\left[(d-3)\Upsilon'(\chi_0)   -\chi_0 \Upsilon''(\chi_0) \right].
\end{align}
Then, let us note the following: if our conjectures \req{conjs} are true, then it follows that there must be a relation between $\Upsilon''(\chi_0)$ and the three-point function parameters $t_2$, $t_4$. More precisely, one can see that this relation would have the form

\begin{equation}
\label{tteap}
a_{(d)} t_2 + b_{(d)} t_4 =\chi_0\frac{\Upsilon''(\chi_0)}{\Upsilon'(\chi_0)}\, ,
\end{equation}
for some constants $a_{(d)}$, $b_{(d)}$ that only depend on the dimension. Based on the results that we currently know, one finds  $a_{(3)}=0$, $b_{(3)}=1/210$, $a_{(5)}=3/20$, $b_{(5)}=23/315$. Now the question is whether these relations are universal, or if they only apply for certain theories.  Let us clarify that \eqref{tteap} might not be universal even if the conjectured relationship between $F_{\mathbb{S}^d_{\varepsilon}}^{(3)}(0)$ and the three-point function charges is. The reason is that, as we said,  \eqref{fee0e} only applies for certain theories. For others, the free energy might be given by a different expression in terms of the gravitational quantities, but nevertheless the relationships between $F_{\mathbb{S}^d_{\varepsilon}}^{(3)}(0)$ and the three-point charges should still hold.

Let us then study the validity of \eqref{tteap} in different dimensions. 
In the case of $d=3$ we have proven that the formula above holds for all theories of the GQT family ---not only for those admitting single-function Taub-NUT solutions. However, we have not checked so far if this result extends more generally to the theories of the Einstein-like type. In any case, this is an interesting result that allows us to compute right away the CFT's three-point function of all holographic GQTGs in $d=3$. 

Next, let us consider the situation in $d=5$. In all examples we have studied, we have seen that all GQTGs that possess single-function Taub-NUT solutions satisfy the relation \eqref{tteap} with the same coefficients ---this is of course necessary if the conjecture \req{5conj3} is true. However, we can study what happens for other GQTGs. Let us consider, as an example, a 6-dimensional action containing quintic terms 
\begin{equation}
S=\frac{1}{16\pi G}\int d^6x \, \sqrt{-g}\left[\frac{20}{L^2}+R+\lambda_5 L^{8} \mathfrak{R}_{(5)}\right]\, ,
\end{equation}
where $\lambda_5$ is a dimensionless parameter and $\mathfrak R_{(5)}$ is a combination of quintic densities given by\footnote{It is not the most general quintic Lagrangian but it is enough for our purposes.}
\begin{align}
\notag
\mathfrak R_{(5)}=&a_1 C_1 Q_1+a_2 C_1 Q_1+a_3 C_1 Q_2+a_4 C_2 R^2+a_5 C_2 Q_1+a_6 C_2 Q_2+a_7 C_3 R^2+a_8 C_3 Q_1\\&\notag+a_9 C_3 Q_2+a_{10} Q_2 R^3+a_{11} Q_1 Q_2 R+a_{12} Q_2^2 R+a_{13} C_5 R^2+a_{14} C_5 Q_1+a_{15}C_5 Q_2\\&+a_{16} C_6 R^2+a_{17} C_6 Q_1+a_{18} C_6 Q_2+a_{19} Q_1 R^3+a_{20} Q_1^2R\, ,
\end{align}
where 
\begin{align}
\notag
Q_1&=R_{ab}R^{ab}\, ,\quad Q_2=R_{abcd}R^{abcd}\, ,\quad C_1=\tensor{R}{_{a}^{c}_{b}^{d}}\tensor{R}{_{c}^{e}_{d}^{f}}\tensor{R}{_{e}^{a}_{f}^{b}}\, ,\quad C_2=\tensor{R}{_{ab}^{cd}}\tensor{R}{_{cd}^{ef}}\tensor{R}{_{ef}^{ab}}\\
C_3&=\tensor{R}{_{a bcd}}\tensor{R}{^{a bc}_{e}}R^{d e}\, ,\quad C_5=\tensor{R}{_{abcd}}\tensor{R}{^{ac}}\tensor{R}{^{bd}}\, ,\quad C_6=R_{a}^{\ b}R_{b}^{\ c}R_{c}^{\ a}\, ,
\end{align}
and where $a_i$ are dimensionless constants. Now, if we restrict ourselves to the subset of these theories that belong to the GQTG class, we find the following constraints 
\begin{align}
a_5&=-\frac{23 a_1}{13}-\frac{17 a_2}{52}-\frac{88 a_4}{13}-\frac{263}{93600}\, ,\\
a_6&=-\frac{25 a_1}{52}-\frac{17 a_3}{52}+\frac{25 a_4}{13}+\frac{607}{336960}\, ,\\
a_8&=\frac{1119 a_1}{130}+\frac{45 a_2}{26}+\frac{462 a_4}{65}-6 a_7+\frac{56773}{4212000}\, ,\\
a_9&=\frac{231 a_1}{52}+\frac{45 a_3}{26}-\frac{231 a_4}{13}-\frac{30373}{1684800}\, ,\\
a_{10}&=-\frac{69 a_1}{260}-\frac{a_3}{26}+\frac{203a_4}{130}+\frac{a_7}{4}+\frac{27637}{50544000}\, ,\\
a_{11}&=\frac{13 a_1}{5}-\frac{9 a_2}{52}+\frac{15 a_3}{26}-\frac{2607a_4}{130}-\frac{11 a_7}{4}-\frac{43859}{4212000}\, ,\\
a_{12}&=-\frac{67 a_1}{104}-\frac{9 a_3}{52}+\frac{67 a_4}{26}+\frac{28811}{10108800}\, ,\\
a_{13}&=-\frac{2 a_1}{5}-\frac{2 a_4}{5}-a_7+\frac{1471}{972000}\, ,\\
a_{14}&=-\frac{322 a_1}{65}-\frac{21 a_2}{13}-\frac{712 a_4}{65}+6 a_7-\frac{270169}{12636000}\, ,\\
a_{15}&=-\frac{76 a_1}{13}-\frac{21 a_3}{13}+\frac{434 a_4}{13}+\frac{162631}{5054400}\, ,\\
a_{16}&=-\frac{221}{194400}\, ,\\
a_{17}&=-\frac{199 a_1}{65}-\frac{11 a_2}{13}-\frac{654 a_4}{65}-3 a_7+\frac{70477}{12636000}\, ,\\
a_{18}&=-\frac{131a_1}{26}-\frac{11a_3}{13}+\frac{327a_4}{13}+\frac{15a_7}{2}+\frac{6113}{5054400}\, ,\\
a_{19}&=-\frac{11 a_1}{130}-\frac{a_2}{26}-\frac{38 a_4}{65}-\frac{1627}{25272000}\, ,\\
a_{20}&=\frac{243a_1}{130}+\frac{15a_2}{26}+\frac{479a_4}{65}+\frac{a_7}{2}+\frac{31553}{12636000}\, ,
\end{align}
so that the remaining free parameters are $a_1,a_2,a_3, a_4, a_7, a_{10}$ and $a_{11}$, together with the overall coupling $\lambda_5$. We have also imposed the standard normalization so that the function $\Upsilon$ reads $\Upsilon(\chi)=1-\chi+\lambda_5 \chi^5$. 
Now, we can compute the energy fluxes as described in the main text and we get the following values for $t_2$ and $t_4$
\begin{equation}
t_2=\frac{40\left(-355+1656\,p\right) \lambda_5 \chi_0^4}{81\left(1-5\lambda_5 \chi_0^4\right)}\, , \quad \quad
t_4=\frac{-1680\,p \lambda_{5} \chi_0^4 }{1-5\lambda_{5}\chi_0^4}\, ,
\end{equation}
where
\begin{equation}
p\equiv \frac{220}{117}-\frac{12150}{13}a_1-\frac{2025}{13}a_2-\frac{810}{13}a_3\, .
\end{equation}
Then, we see that 
\begin{equation}
\frac{3}{20} t_2 +\frac{23}{315} t_4 -\chi_0\frac{\Upsilon''(\chi_0)}{\Upsilon'(\chi_0)}=-\frac{10\left(17+2p\right) \lambda_5 \chi_0^4}{27\left(1-5\lambda_5 \chi_0^4\right)}\, ,
\end{equation}
which is in general nonzero,\footnote{For $p=-17/2$ the relationship \eqref{tteap} holds, but the corresponding theory (actually, set of theories) does not allow for single-function Taub-NUT solutions. Therefore the implication only holds in one direction: if a theory possesses single-function Taub-NUT solutions, then it satisfies \eqref{tteap}, but the converse is not true.} hence implying that \eqref{tteap} does not apply universally to all GQTGs in $d=5$.  We expect the same behaviour to happen in higher dimensions. 
Therefore, the conclusion is that (if the conjectures in \req{conjs} are true) the relation \eqref{tteap} holds for all $d=3$ and $d=5$ GQT theories possessing single-function Taub-NUT solutions, but it is not satisfied by all GQT gravities (except in $d=3$) or by general Einstein-like theories. 
Let us also mention that in the case of even $d$ we do not have any argument to support the existence of relations of the form \eqref{tteap}. In fact, in $d=4$ we have checked that there is no way to accommodate the coefficients $a_{(4)}$, $b_{(4)}$ so that \eqref{tteap} is satisfied simultaneously by GB gravity and QT and GQT gravities.

\section{Additional details for quintic theories}
\label{moreQuintic}

Here we collect some results and discussion that is too cumbersome for the main text. The complete basis of invariants used to construct the quintic densities is
\begin{align}
q_1 &= R_{ab} R^{ab} \, , \quad q_2 = R_{abcd} R^{abcd}
\nonumber\\
C_1 &= \tensor{R}{_a ^b _c ^d} \tensor{R}{_b ^e _d ^f} \tensor{R}{_e ^a _f ^c} \, , \quad C_2 = \tensor{R}{_a _b ^c ^d} \tensor{R}{_c _d ^e ^f} \tensor{R}{_e _f ^a ^b}  \, , \quad C_3 = \tensor{R}{_a _b _c _d} \tensor{R}{^a ^b ^c _e} \tensor{R}{^d ^e} \, , \quad C_5 = \tensor{R}{_a _b _c _d} \tensor{R}{^a ^c} \tensor{R}{^b ^d}\, ,
\nonumber\\
C_6 &= \tensor{R}{_a ^b} \tensor{R}{_b ^c} \tensor{R}{_c ^a} \, ,
\nonumber\\
Q_1 &= \tensor{R}{^\mu ^\nu ^\rho ^\sigma} \tensor{R}{_\mu ^\delta _\rho ^\gamma} \tensor{R}{_\delta ^\chi _\nu ^\xi}  \tensor{R}{_\gamma _\chi _\sigma _\xi} \, , \quad Q_2 =  \tensor{R}{^\mu ^\nu ^\rho ^\sigma} \tensor{R}{_\mu ^\delta _\rho ^\gamma}\tensor{R}{_\delta ^\chi _\gamma ^\xi} \tensor{R}{_\nu _\chi _\sigma _\xi}  \, , Q_3 = \tensor{R}{^\mu ^\nu ^\rho ^\sigma} \tensor{R}{_\mu _\nu ^\delta ^\gamma} \tensor{R}{_\rho ^\chi _\delta ^\xi}\tensor{R}{_\sigma _\chi _\gamma _\xi}
\nonumber\\
Q_4 &= \tensor{R}{^\mu ^\nu ^\rho ^\sigma}\tensor{R}{_\mu _\nu ^\delta ^\gamma}\tensor{R}{_\rho _\delta ^\chi ^\xi}\tensor{R}{_\sigma _\gamma _\chi _\xi} \, , \quad Q_5 = \tensor{R}{^\mu ^\nu ^\rho ^\sigma}\tensor{R}{_\mu _\nu ^\delta ^\gamma}\tensor{R}{_\delta _\gamma ^\chi ^\xi}\tensor{R}{_\rho _\sigma _\chi _\xi} \, , \quad Q_6 = \tensor{R}{^\mu ^\nu ^\rho ^\sigma} \tensor{R}{_\mu _\nu _\rho ^\delta} \tensor{R}{_\gamma _\xi _\chi _\sigma} \tensor{R}{^\gamma ^\xi ^\chi _\delta} \, , 
\nonumber\\
Q_8 &= \tensor{R}{^\mu ^\nu} \tensor{R}{^\rho ^\sigma ^\delta ^\gamma}\tensor{R}{_\rho ^\xi _\delta _\mu}\tensor{R}{_\sigma _\xi _\gamma _\nu} \, , \quad Q_9 = \tensor{R}{^\mu ^\nu}\tensor{R}{^\rho ^\sigma ^\delta ^\gamma}\tensor{R}{_\rho _\sigma ^\xi _\mu}\tensor{R}{_\delta _\gamma _\xi _\nu} \, , \quad Q_{10} = \tensor{R}{^\mu ^\nu}\tensor{R}{_\mu ^\rho _\nu ^\sigma}\tensor{R}{_\delta _\gamma _\xi _\rho}\tensor{R}{^\delta ^\gamma ^\xi _\sigma}\, , 
\nonumber\\
Q_{13} &= \tensor{R}{^\mu ^\nu}\tensor{R}{^\rho ^\sigma}\tensor{R}{^\delta _\mu ^\gamma _\rho}\tensor{R}{_\delta _\nu _\gamma _\sigma} \, , \quad Q_{14} = \tensor{R}{^\mu ^\nu}\tensor{R}{^\rho ^\sigma}\tensor{R}{^\delta _\mu ^\gamma _\nu}\tensor{R}{_\delta _\rho _\gamma _\sigma} \, ,  \quad Q_{15} = \tensor{R}{^\mu ^\nu}\tensor{R}{^\rho ^\sigma}\tensor{R}{^\delta ^\gamma _\mu _\rho}\tensor{R}{_\delta _\gamma _\nu _\sigma} \, , \quad
\nonumber\\
Q_{16} &= \tensor{R}{^\mu ^\nu}\tensor{R}{_\nu ^\rho}\tensor{R}{^\sigma ^\delta ^\gamma _\mu}\tensor{R}{_\sigma _\delta _\gamma _\rho} \, \quad  Q_{22} = \tensor{R}{_\mu ^\nu}\tensor{R}{_\nu ^\rho}\tensor{R}{_\rho ^\sigma}\tensor{R}{_\sigma ^\mu} 
\nonumber\\
H_1 &= \tensor{R}{_a _b ^c ^d}\tensor{R}{_c _d ^e ^f}\tensor{R}{_e _f ^m ^n}\tensor{R}{_m _n ^r ^s}\tensor{R}{_r _s ^a ^b} \, , \quad H_2 = \tensor{R}{_c _e ^a ^b}\tensor{R}{_a _g ^c ^d}\tensor{R}{_b _i ^e ^f}\tensor{R}{_f _j ^g ^h}\tensor{R}{_d _h ^i ^j}
\nonumber\\
H_3 &= \tensor{R}{_c _e ^a ^b}\tensor{R}{_a _f ^c ^d}\tensor{R}{_g _i ^e ^f}\tensor{R}{_b _j ^g ^h}\tensor{R}{_d _h ^i ^j} \, , \quad H_4 = \tensor{R}{_c _d ^a ^b} \tensor{R}{_e _g ^c ^d}\tensor{R}{_a _i ^e ^f}\tensor{R}{_f _j ^g ^h}\tensor{R}{_b _h ^i ^j}
\nonumber\\
H_5 &= \tensor{R}{_b ^a}\tensor{R}{_d _f ^b ^c}\tensor{R}{_g _h ^d ^e}\tensor{R}{_e _i ^f ^g}\tensor{R}{_a _c ^h ^i} \, , \quad H_6 = \tensor{R}{_b ^a}\tensor{R}{_d _f ^b ^c}\tensor{R}{_a _h ^d ^e}\tensor{R}{_e _i ^f ^g}\tensor{R}{_c _g ^h ^i}
\nonumber\\
H_7 &= \tensor{R}{_b ^a}\tensor{R}{_d _f ^b ^c}\tensor{R}{_a _c ^d ^e}\tensor{R}{_h _i ^f ^g}\tensor{R}{_e _g ^h ^i} \, , \quad H_8 = \tensor{R}{_b ^a}\tensor{R}{_d _e ^b ^c}\tensor{R}{_c _f ^d ^e}\tensor{R}{_h _i ^f ^g}\tensor{R}{_a _g ^h ^i}
\nonumber\\
H_9 &= \tensor{R}{_b ^a}\tensor{R}{_a _d ^b ^c}\tensor{R}{_f _h ^d ^e}\tensor{R}{_c _i ^f ^g}\tensor{R}{_e _g ^h ^i}\, , \quad H_{10} = \tensor{R}{_c ^a}\tensor{R}{_e ^b}\tensor{R}{_a _f ^c ^d}\tensor{R}{_g _h ^e ^f}\tensor{R}{_b _d ^g ^h}
\nonumber\\
H_{11} &= \tensor{R}{_c ^a}\tensor{R}{_d ^b}\tensor{R}{_e _g ^c ^d}\tensor{R}{_a _h ^e ^f}\tensor{R}{_b _f ^g ^h} \, , \quad H_{12} = \tensor{R}{_c ^a}\tensor{R}{_d ^b}\tensor{R}{_e _f ^c ^d}\tensor{R}{_g _h ^e ^f}\tensor{R}{_a _b ^g ^h}
\nonumber\\
H_{13} &= \tensor{R}{_c ^a}\tensor{R}{_d ^b}\tensor{R}{_a _e ^c ^d}\tensor{R}{_g _h ^e ^f}\tensor{R}{_b _f ^g ^h} \, , \quad H_{14} = \tensor{R}{_c ^a}\tensor{R}{_d ^b}\tensor{R}{_a _b ^c ^d}\tensor{R}{_g _h ^e ^f}\tensor{R}{_e _f ^g ^h}
\nonumber\\
H_{15} &= \tensor{R}{_b ^a}\tensor{R}{_c ^b}\tensor{R}{_e _g ^c ^d}\tensor{R}{_a _h ^e ^f}\tensor{R}{_d _f ^g ^h} \, , \quad H_{16} = \tensor{R}{_b ^a}\tensor{R}{_c ^b}\tensor{R}{_e _f ^c ^d}\tensor{R}{_g _h ^e ^f}\tensor{R}{_a _d ^g ^h}
\nonumber\\
H_{17} &= \tensor{R}{_b ^a}\tensor{R}{_c ^b}\tensor{R}{_a _e ^c ^d}\tensor{R}{_g _h ^e ^f}\tensor{R}{_d _f ^g ^h} \, , \quad H_{18}  = \tensor{R}{_b ^a}\tensor{R}{_d ^b}\tensor{R}{_f ^c}\tensor{R}{_c _g ^d ^e}\tensor{R}{_a _e ^f ^g}
\nonumber\\
H_{19} &= \tensor{R}{_b ^a}\tensor{R}{_d ^b}\tensor{R}{_f ^c}\tensor{R}{_a _g ^d ^e}\tensor{R}{_c _e ^f ^g}
\end{align}
From this basis, we identified two distinct GQT terms that admit single-function Taub-NUT solutions:
\begin{align}
\mathcal{Q}_1 &=   -{\frac {7920397876\,H_{3}}{7007602661}}+{\frac {458341466043\,H_{6}}{
175190066525}}+{\frac {924457527463\,H_{9}}{1576710598725}}-{\frac {
11148655827131\,H_{10}}{11352316310820}}
\nonumber\\
&+{\frac {5069081933699\,H_{17}
}{18920527184700}}+{\frac {33895725083749\,H_{19}}{102170846797380}}-{
\frac {34062780623603\,Rq_{1}\,q_{2}}{16347335487580800}}
\nonumber\\
&+{\frac {
49841316682649\,q_{1}\,C_{2}}{3632741219462400}}+{\frac {
649726495852789\,q_{1}\,C_{3}}{5449111829193600}}-{\frac {
5767082868853271\,q_{1}\,C_{5}}{24521003231371200}}
\nonumber\\
&-{\frac {
5032299709877453\,q_{1}\,C_{6}}{24521003231371200}}-{\frac {
233963380511\,q_{2}\,C_{2}}{24218274796416}}-{\frac {47730869615569\,q
_{2}\,C_{6}}{908185304865600}}
\nonumber\\
&+{\frac {642494777099\,q_{2}\,C_{3}}{
8409123193200}}+{\frac {20850012672071\,q_{2}\,C_{5}}{544911182919360}
}-{\frac {3431709507561\,RQ_{1}}{11212164257600}}
\nonumber\\
&-{\frac {
125078101388107\,RQ_{10}}{681138978649200}}-{\frac {51784130212603\,RQ
_{13}}{681138978649200}}+{\frac {43909166340607\,RQ_{14}}{
6130250807842800}}
\nonumber\\
&+{\frac {6047423933369\,RQ_{2}}{100909478318400}}-{
\frac {69576046485937\,RQ_{15}}{2724555914596800}}+{\frac {
3688080828391\,RQ_{6}}{201818956636800}}-{\frac {85555281053\,RQ_{5}}{
10090947831840}}
\nonumber\\
&+{\frac {10482876107767\,RQ_{9}}{60545686991040}}-{
\frac {416697046093847\,{R}^{2}C_{2}}{21796447316774400}}+{\frac {
1111035771576389\,{R}^{2}C_{3}}{32694670975161600}}
\nonumber\\
&+{\frac {
334531709737151\,{R}^{2}C_{5}}{16347335487580800}}+{\frac {
2861130671335589\,{R}^{2}C_{6}}{147126019388227200}}-{\frac {
59879809392517\,q_{1}\,C_{1}}{908185304865600}}
\nonumber\\
&+{\frac {538470703453\,
q_{2}\,C_{1}}{6054568699104}}+{\frac {9498199270340447\,R{q_{1}}^{2}}{
98084012925484800}}-{\frac {2359210333399\,R{q_{2}}^{2}}{
201818956636800}}
\nonumber\\
&-{\frac {66114135677249\,{R}^{2}C_{1}}{
5449111829193600}}-{\frac {1370761460727169\,{R}^{3}q_{1}}{
98084012925484800}}-{\frac {17884717886671\,{R}^{3}q_{2}}{
32694670975161600}}
\nonumber\\
&+{\frac {441362262685967\,{R}^{5}}{
1177008155105817600}}
\end{align}
\begin{align}
\mathcal{Q}_2 &= -{\frac {2333677396\,H_{3}}{2696708431}}+{\frac {167522524618\,H_{6}}{
67417710775}}+{\frac {827734758038\,H_{9}}{606759396975}}
\nonumber\\
&+{\frac {
1580949633097\,H_{10}}{2184333829110}}+{\frac {593519819837\,H_{17}}{
3640556381850}}-{\frac {54396973475063\,H_{19}}{19659004461990}}
\nonumber\\
&+{
\frac {226426936847281\,Rq_{1}\,q_{2}}{3145440713918400}}+{\frac {
1237975085831\,q_{1}\,C_{2}}{25888400937600}}-{\frac {160999722984743
\,q_{1}\,C_{3}}{1048480237972800}}
\nonumber\\
&+{\frac {363222093684877\,q_{1}\,C_{
5}}{4718161070877600}}+{\frac {1536726748058911\,q_{1}\,C_{6}}{
4718161070877600}}-{\frac {180016228721\,q_{2}\,C_{2}}{13979736506304}
}
\nonumber\\
&-{\frac {6450313084397\,q_{2}\,C_{6}}{174746706328800}}+{\frac {
342806058161\,q_{2}\,C_{3}}{4854075175800}}-{\frac {6045492677197\,q_{
2}\,C_{5}}{104848023797280}}
\nonumber\\
&-{\frac {14884840514911\,RQ_{1}}{
58248902109600}}-{\frac {48630288598591\,RQ_{10}}{131060029746600}}+{
\frac {3021059012761\,RQ_{13}}{131060029746600}}
\nonumber\\
&+{\frac {
983210397728611\,RQ_{14}}{1179540267719400}}-{\frac {342538077427\,RQ_
{2}}{174746706328800}}-{\frac {4146157105207\,RQ_{15}}{174746706328800
}}
\nonumber\\
&+{\frac {6534697440049\,RQ_{6}}{116497804219200}}-{\frac {
119311421347\,RQ_{5}}{34949341265760}}+{\frac {815141983411\,RQ_{9}}{
11649780421920}}
\nonumber\\
&-{\frac {5653397143411\,{R}^{2}C_{2}}{4193920951891200
}}+{\frac {40312570800617\,{R}^{2}C_{3}}{6290881427836800}}-{\frac {
307449586546597\,{R}^{2}C_{5}}{3145440713918400}}
\nonumber\\
&-{\frac {
184522325982583\,{R}^{2}C_{6}}{28308966425265600}}-{\frac {
10757768133107\,q_{1}\,C_{1}}{58248902109600}}+{\frac {349701812335\,q
_{2}\,C_{1}}{3494934126576}}
\nonumber\\
&-{\frac {2371496498259589\,R{q_{1}}^{2}}{
18872644283510400}}-{\frac {1777541084311\,R{q_{2}}^{2}}{
116497804219200}}-{\frac {28399488598837\,{R}^{2}C_{1}}{
1048480237972800}}
\nonumber\\
&+{\frac {519486669787883\,{R}^{3}q_{1}}{
18872644283510400}}-{\frac {26848314588043\,{R}^{3}q_{2}}{
6290881427836800}}-{\frac {354830859472789\,{R}^{5}}{
226471731402124800}}
\end{align}
When evaluated on static geometries the field equations for these theories (including also the usual Einstein-Hilbert term) read
\be  
\frac{4 r^5 (f-1)}{L^2} + \mu \mathcal{F}_{\mathcal{Q}_1} + \beta \mathcal{F}_{\mathcal{Q}_2} = \frac{16 \pi G M}{\Omega_k} \, ,
\ee
where we have, as in Eq.~\eqref{eq:hypBH3d}, defined
\be 
V(r) \equiv \frac{r^2 f}{L^2} + k \, ,
\ee
and
\begin{align}
\mathcal{F}_{\mathcal{Q}_1} =& \frac {22335140604\,{r}^{5}}{7007602661\,{L}^{10}} \bigg[ {r}^{2}ff'^{2} \left( {L}^{2}k+f{r}^{2} \right) f'' +\frac{1}{4} {L}^{2}f'^{4}k{r}^{2}+ \left( 3 k{L}^{2}rf+\frac{7}{3}{f}^{2}{r}^{3}
 \right) f'^{3}
 \nonumber\\
 &+{\frac {7007602661\,{f}^{5}}{5583785151}}
 \bigg]
\end{align}


\begin{align}
\mathcal{F}_{\mathcal{Q}_2} =& \frac {3722523434 {r}^{5}}{2696708431 {L}^{10}} \bigg[ {r}^{2}f'^{2} \left( rf'+6 f \right)  \left( {L}^{2}k+{r}^{2}f
 \right) f''-\frac{1}{5} f'^{5}{r}^{5}+ 2\left( 2 {r}^{2}{L}^{2
}k+ f{r}^{4} \right) f'^{4} 
\nonumber\\
&+ 2 \left( 9 k{L}^{2}rf+7 {f}^{
2}{r}^{3} \right) f'^{3}+{\frac {5393416862 {f}^{5}}{1861261717}} \bigg]
\end{align}
%
It is evident from these expressions that $\mathcal{Q}_1$ and $\mathcal{Q}_2$ are \textit{distinct} GQT theories.

\bibliographystyle{JHEP}
\bibliography{Gravities}

\end{document}